\documentclass[
 reprint,
 amsmath,amssymb,fleqn,
 aip,
 jcp,
]{revtex4-1}

\usepackage[english]{babel}
\usepackage[utf8x]{inputenc}
\usepackage{amssymb}
\usepackage{amsmath}
\usepackage{amsfonts}
\usepackage[bbgreekl]{mathbbol}
\usepackage{graphicx}
\usepackage{color}
\usepackage{hyperref}

\newcommand{\bra}{\langle}
\newcommand{\ket}{\rangle}
\newcommand{\QQ}{\mathbf{Q}}

\newcommand{\ee}{\mathrm{e}}

\title{Efficient simulation of the photodynamics of dihalogens in solid matrices using the G-MCTDH method. \\ 
Quantum dynamics and spectroscopy of the embedded $\mathrm{I_2 Kr_{18}}$ cluster.}
\begin{document}

\title{Quantum dynamics and spectroscopy of dihalogens in solid matrices.\\ 
I.\ Efficient simulation of the photodynamics of the
embedded $\mathrm{I_2 Kr_{18}}$ cluster using the G-MCTDH method.}
%Quantum dynamics and spectroscopy of the embedded $\mathrm{I_2 Kr_{18}}$ cluster.}
\author{David Picconi}
\email{picconi@chemie.uni-frankfurt.de}
\affiliation{Institute of Physical and Theoretical Chemistry, Goethe University Frankfurt, 
 Max-von-Laue-Stra{\ss}e 7, D-60438 Frankfurt am Main, Germany}
\author{Jeffrey A. Cina}
\affiliation{Department of Chemistry and Biochemistry, and Oregon Center for Optical, Molecular, and Quantum Science, University of Oregon, Eugene, Oregon 97403, USA}
\author{Irene Burghardt}
\affiliation{Institute of Physical and Theoretical Chemistry, Goethe University Frankfurt,  Max-von-Laue-Stra{\ss}e 7, D-60438 Frankfurt am Main, Germany}

\date{\today}

\begin{abstract}
The molecular dynamics following the electronic $B\ ^3\Pi_u\left(0^+\right) \longleftarrow X\ ^1\Sigma_g^+$
photoexcitation of the iodine molecule embedded in solid krypton are studied
quantum mechanically using the Gaussian variant of the multiconfigurational
time-dependent Hartree method (G-MCTDH). The accuracy of the Gaussian
wave packet approximation is validated against numerically exact MCTDH
simulations for a fully anharmonic seven-dimensional model of the $\mathrm{I_2
Kr_{18}}$ cluster in a crystal Kr cage. The linear absorption spectrum,
time-evolving vibrational probability densities, and $\mathrm{I_2}$ energy expectation
value are accurately reproduced by the numerically efficient G-MCTDH approach.
The reduced density matrix of the chromophore is analyzed in the
coordinate, Wigner and energy representations, so as to obtain a multifaceted dynamical view of the guest-host interactions.
Vibrational coherences
extending over the bond distance range $ 2.7\,\mbox{\AA} < R_\mathrm{I-I} < 4.0\,\mbox{\AA}$
are found to survive for several vibrational periods, despite extensive
dissipation. The present results prepare the ground for the simulation of
time-resolved coherent Raman spectroscopy of the $\mathrm{I_2}$-krypton system
addressed in a companion paper.
%This kind of coherences can be monitored by time-resolved coherent Raman
%spectroscopy performed on a superposition of wave packets created on the $B$
%state potential energy surface by a pair of pump pulses. A theoretical
%description of such experiment is developed, which elucidates the connection
%between the nonlinear signal and the wave packet coherence, and provides an
%effective means to simulate the spectra with a minimum number of quantum
%dynamical propagations. The validated G-MCTDH method is used to simulate and
%interpret the time-resolved coherent Raman signals for two selected initial
%superpositions; the mechanisms which drive decoherence are related to the
%details of the solvent motion and depend on the specific initial superposition
%in a non-trivial way.
\end{abstract}

\maketitle

\section{Introduction}
The fundamental aspects of the molecular dynamics of chromophores embedded in
a complex environment are the subject of extensive studies in contemporary
chemical physics.\cite{nitzan-book,PN06,RLV13} A number of molecular processes
of interest in biochemistry or material science are typically initiated by the
interaction with light which induces an electronic excitation. The absorption
of one photon prepares the chromophore in a coherent superposition of
vibrational and electronic states which is subject to dynamical evolution. The
presence of the environment induces a loss of coherence, whose elementary
mechanisms depend on the interactions of the chromophore with its surroundings
in a complicated way.\cite{J18} The detailed description of quantum coherences
is essential to understand the mode of operation of photosynthetic biological
systems\cite{ECRAMCBF07,PHFCHWBE10,SSS12} or functional
materials.\cite{CS09,BSS17} These details can be unraveled by today's sophisticated
techniques of nonlinear spectroscopy,\cite{DSM16} which provide third- or
higher-order-spectra\cite{C01,DFKHWB18} whose interpretation requires advanced
theoretical analysis.\cite{mukamel,GED09}

Furthermore, the understanding of vibrationally coherent phenomena is
instrumental to our ability of controlling chemical reactivity using laser
pulses.\cite{ABBBKSSG98,D01} Vibrational coherences were observed
experimentally in biopolyenes,\cite{BM14} including the visual pigment
retinal,\cite{JHMPEM15} transition metal complexes,\cite{MCAGVTC18} molecular
aggregates\cite{SY11,BVA14} and small molecules embedded in solid
matrices at cryogenic temperatures.\cite{CBDGGO00,GAEC02,GBFKS07} Diatomic halogen molecules in rare-gas
crystals have long been investigated by spectroscopists, as they allow a
highly detailed scrutiny of a number of processes prototypical for condensed
phase photodynamics,\cite{AS99,GBFKS07,kuehn-woeste-book} such as bond elongation and geminate
recombination upon molecule-cage collision,\cite{KBGSS04,BZKA04} nonadiabatic
dynamics\cite{BC97,BGS04,BZKA04} and energy dissipation to the
solvent.\cite{BGDS02} Time-resolved coherent Raman scattering experiments
performed on the $\mathrm{I_2}$ chromophore in solid Kr allowed to follow the
evolution of the vibrational coherence for several picoseconds, despite
extensive dissipation,\cite{SKFA05} and the dependence of the decoherence
mechanism on the preparation pulses was demonstrated.\cite{SA11} This and
related optical experiments\cite{GBS03,KAMAP05,GSA13} are based on the
controlled realization of a quantum mechanical superposition of wave packets
interacting with the cage.

The present paper focuses on the simulation and theoretical interpretation of
such chromophore-environment states. A quantum mechanical treatment of all relevant nuclear motion is necessary for a rigorous description of (i) coherent time-resolved spectroscopies,\cite{PCB18B} and (ii) experiments that are carried out at temperatures much lower than that corresponding to the Debye frequency of a krypton crystal, so that the zero-point motion of the lattice phonons cannot be ignored.
Due to the specificities of the interaction, quantum dynamical simulations must explicitly
include the degrees of freedom of both the chromophore and the rare-gas bath.
In the last decades, several such methods for high-dimensional quantum wave
packet propagation have been developed and tested. Among them, the
multi-configurational time-dependent Hartree (MCTDH) approach\cite{BJWM00} and
the related multi-layer variant\cite{WT03} have been applied to
system-environment problems whose dimensionality ranges from dozens to
hundreds degrees of freedom.\cite{mctdh-book} In a parallel effort, more
approximate quantum dynamical treatments of the bath dynamics have been
developed, based on the use of Gaussian wave packets. Here, a small number of
coordinates (system modes) directly correlated with the electronic excitation,
or involved in a reaction mechanism, are described accurately, whereas a
Gaussian ansatz is used for the secondary degrees of freedom. This type of
approach includes, in particular, the Gaussian-based MCTDH variant
(G-MCTDH),\cite{BMC99} in which the standard single-particle functions are
replaced for the bath modes by Gaussian wave packets and the propagation is
fully variational. A similar approach is the local coherent state
approximation (LCSA) of Martinazzo et al.,\cite{MNST06} in which the
vibrational basis is replaced by a discrete variable representation (DVR).
More recently, Kovac and Cina\cite{KC17}, described the photodynamics of a
$\mathrm{I_2 Kr_6}$ model cluster using the fixed vibrational basis/Gaussian
bath theory (FVB/GB),\cite{CC07,CCC11,CC14} in which the wavefunction for the
system coordinate (the I--I stretch) is represented on a conventional basis of
vibrational eigenstates, and each level is accompanied by a Gaussian wave
packet describing the bath. In all of these approaches, the propagation is
fully based on the time-dependent variational principle.

In the present work, we adopt the G-MCTDH approach for the study of a larger
$\mathrm{I_2 Kr_{18}}$ cluster, related to the model of
Ref. \onlinecite{KC17}. In further work, we aim to systematically develop
Gaussian wave packet based approaches to efficiently simulate experiments of
nonlinear spectroscopy, in which several calculations must be launched for
different time separations between the pulses, carrier frequencies,
etc.\cite{GED09} The G-MCTDH method has so far been successfully tested for
standard model potentials,\cite{WB03} harmonic molecules\cite{BGW08} and
anharmonic systems coupled to a harmonic bath.\cite{BNW03}
No computational applications have been performed, however, 
for the dynamics of molecules
embedded in an environment that interacts via anharmonic (van der Waals)
forces through specific solvent modes. 
Even though, in principle, a general force field can be always mapped to the Caldeira-Leggett
model of independent harmonic oscillators,\cite{CL81,GIK16}
the atomistic viewpoint may be more natural, and 
the possibility of visualizing selected bath modes may be informative about the molecular
mechanisms of dissipation and decoherence. Additional developments are also
needed concerning spectroscopic simulations. Despite the fact that MCTDH and related methods
have a huge potential, they are actually rarely applied to the computation of
nonlinear signals and to the analysis of spectroscopic signatures of the
system-bath interactions.

\begin{figure}[b!]
\centering
\includegraphics[scale=0.7]{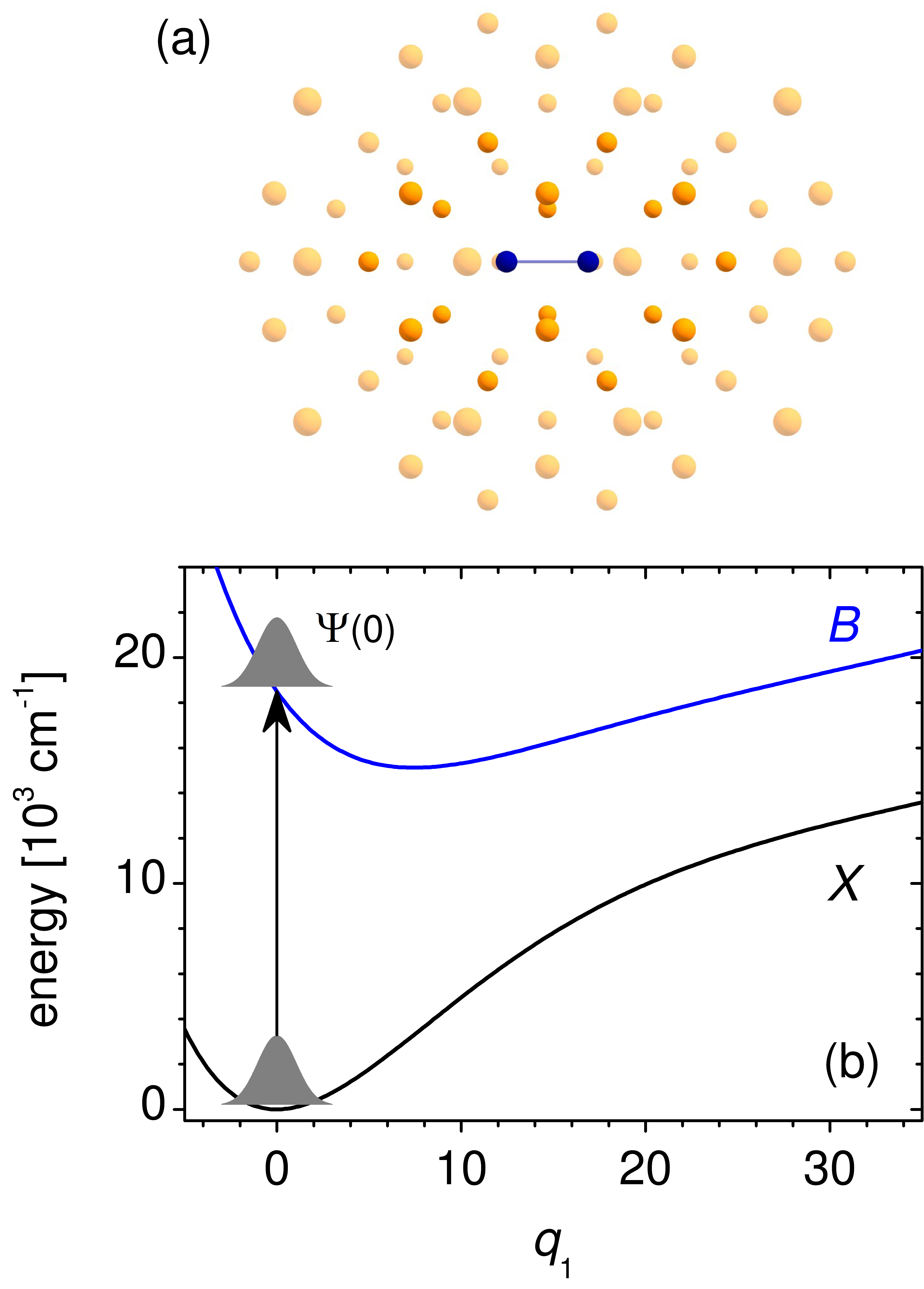}
\caption{(a)  Optimized structure of the $\mathrm{I_2 Kr_{18}}$ cluster embedded into a frozen cage of 54 Kr atoms (shown as transparent). (b) One-dimensional potential energy cuts of the electronic states $X$ and $B$ of the $\mathrm{I_2 Kr_{18}}$ cluster along the normal mode $q_1$, which largely corresponds to the I--I stretch. $\Psi(0)$ is the initial seven-dimensional Gaussian wave packet used in the MCTDH and G-MCTDH(I-IV) calculations (see Table \ref{tab: MCTDH details}) and the black arrow signifies the $B \longleftarrow X$ Franck-Condon excitation.}
\label{fig: PECs}
\end{figure}

The purpose of the present paper is therefore to validate the quality of the
G-MCTDH description of the anharmonic system-bath interaction for the photoinduced coherent dynamics
in a $\mathrm{I_2 Kr_{18}}$ cluster, by comparing with numerically exact
results obtained by the computationally more demanding MCTDH method.
To this end, a reduced dimensionality anharmonic Hamiltonian is set up for the
electronic states $X \ ^1\Sigma_g^+$ and $B \ ^3\Pi_u(0^+)$ of the iodine molecule embedded in crystal
krypton (see Fig. \ref{fig: PECs}(b) for a sketch of the $X$ and $B$
surfaces).\cite{GBFKS07} The model includes the system mode (i. e. the I--I
stretch) and a minimal number of six bath modes, so that numerically exact quantum
dynamical simulations can be performed. The second purpose is the
identification of the main aspects of the dynamics of the $\mathrm{I_2:Kr}$
system, following the $B \leftarrow X$ electronic photoexcitation. The focus
is on the evolution of quantum mechanical coherences and how they can be
monitored by measurements of observables \lq projected' on the chromophore
coordinate. This is facilitated by the analysis of the system reduced density
matrices simultaneously in different representations: The coordinate picture
(which provides the spatial extension of vibrational coherences), the Wigner
function (which allows to visualize the quasi-classical phase space motion of
the vibrational wave packet) and the energy representation (which gives
information about dissipation).

The present study prepares the ground for the second part of the
investigation, reported in the companion paper,\cite{PCB18B} where the simulation of time-resolved
coherent Raman spectroscopy of $\mathrm{I_2}$ embedded in krypton is addressed.

%Once the validity of the G-MCTDH method has been proved, its ability in
%simulating nonlinear spectroscopic signals has to be checked. In this respect,
%an ideal test-bed are the time-resolved four-wave mixing experiments of
%Apkarian and coworkers,\cite{SKFA05,SA11} which are aimed at monitoring the
%evolution of the vibrational coherence between two wave packets created by a
%pair of pump pulses. The third goal of this paper is to show that the G-MCTDH
%method is able to predict accurately the time evolution of such vibrational
%coherence and, as a consequence, the nonlinear spectrum. To this end, a
%theoretical analysis of the spectra is developed to simulate the signals using
%a minimum amount of quantum dynamical calculations, and to establish a
%quantitative connection between the vibrational coherence and the spectra.
%Coherent Raman spectra are then calculated for different initial wave packet
%pairs, which mimic hypothetical superpositions created by a sequence of two
%pump pulses. The nontrivial features of the theoretical and experimental
%signals are then compared in detail.

The manuscript is organized as follows. In Sect. \ref{sec: theory} the
formulation of the G-MCTDH ansatz and the equations of motions are reported,
in order to make the manuscript self-contained, and the construction of the
Hamiltonian of the $\mathrm{I_2 Kr_{18}}$ cluster is explained; in Sect.
\ref{sec: Comp. Det.} the computational details of the calculations are given;
Sect. \ref{sec: mctdh vs gmctdh} reports the comparison between the MCTDH and
G-MCTDH simulations for the $B \leftarrow X$ excitation of the $\mathrm{I_2
Kr_{18}}$ cluster, and the analysis of observables and reduced density
matrices. Finally, Sec.\ \ref{sec: Conclusion} concludes. 
%in Sect. \ref{sec: cat states} the approximate method to evaluate
%time-resolved coherent Raman scattering signals is developed and applied to
%two different initial wave packet superposition, the spectral signatures of
%the dissipation mechanism are identified and associated to specific vibrations
%involving the Kr atoms of the first solvation shell; Sect. \ref{sec:
%Conclusion} summarizes the results and discusses future prospects.

\section{Theory}
\label{sec: theory}
\subsection{The G-MCTDH method}
In the G-MCTDH method\cite{BMC99,WB03,BGW08} the nuclear time-dependent wavefunction is expressed in terms of the coordinates of $P$ \lq particles' $(\QQ_1,...,\QQ_P)$, each of them representing a subset of the \lq physical' degrees of freedom, $\QQ_\kappa = \left(q_{\kappa 1}, q_{\kappa 2}, ..., q_{\kappa n_\kappa} \right)$. The wavefunction has the multi-configurational form
\begin{equation}
\hspace{-0.5cm} \Psi(\QQ_1,...,\QQ_P) = \sum_J A_J(t) \Phi_J(\QQ_1,...,\QQ_P,t)  \label{eq: GMCTDH_wf}
\end{equation}
where each configuration corresponds to a Hartree product of time-dependent single-particle functions (SPFs),
\begin{equation}
\hspace{-0.5cm} \Phi_J(\QQ_1,...,\QQ_P,t) = \prod_{\kappa = 1}^M \varphi_{j_\kappa}^{(\kappa)}(\QQ_\kappa,t) \prod_{\kappa = M + 1}^{P} g_{j_\kappa}^{(\kappa)}(\QQ_\kappa,t) \ .
\end{equation}
$M$ \lq primary' particles are described using general SPFs $\varphi_j^{(\kappa)}$ which are represented on $n_\kappa-$dimensional grids, as in the standard MCTDH method. The SPFs of the remaining $P - M$ \lq secondary' particles are Gaussian wave packets (GWPs) of general form 
\begin{eqnarray}
\hspace{-0.5cm} g_j^{(\kappa)}(\QQ_\kappa,t) &  \nonumber \\
 & \hspace{-1.5cm} =  \exp\left[\QQ_\kappa \cdot \mathbf{a}_j^{(\kappa)}(t) \cdot \QQ_\kappa + \boldsymbol{\xi}_j^{(\kappa)}(t) \cdot \QQ_\kappa + \eta_j^{(\kappa)}(t) \right] \ , \nonumber \\
 & 
\end{eqnarray}
and parametrized by the complex numbers $\boldsymbol{\Lambda}_j^{(\kappa)} = \left\{ \mathbf{a}_j^{(\kappa)}, \boldsymbol{\xi}_j^{(\kappa)}, \eta_j^{(\kappa)}.  \right\} $. In the present applications the so-called \lq frozen' Gaussians are used, in which the width matrix $\mathbf{a}_j^{(\kappa)}$ is chosen to be real, diagonal and constant in time, so that the GWPs are written as
\begin{equation}
\hspace{-0.5cm} g_j^{(\kappa)}(\QQ_\kappa,t) = \ee^{\eta_j^{(\kappa)}} \prod_{\alpha = 1}^{n_\kappa} \exp\left[ - a_{j\alpha}^{(\kappa)} q_{\kappa \alpha}^2 + \xi_{j\alpha}^{(\kappa)}(t) q_{\kappa \alpha} \right] \ ;
\end{equation}
moreover the exponents $\eta_j^{(\kappa)}$ are chosen to be real and to provide normalized GWPs at each time.

The equations for the time evolution of the parameters are obtained by applying the Dirac-Frenkel time-dependent variational principle\cite{MMC90}, $\left\bra \delta \Psi \left | \hat{H} - i\hbar \partial / \partial t \right| \Psi \right\ket = 0$, to the ansatz of Eq. (\ref{eq: GMCTDH_wf}) and by removing redundancies using the constraints $\left\bra \varphi_j^{(\kappa)}(t) \left| \dot{\varphi}_l^{(\kappa)}(t) \right\ket \right. = 0$ for $\kappa = 1, ..., M$. In this way three sets of coupled equations of motion are derived:\cite{BMC99,BGW08}
\begin{itemize}
\item The time evolution of the coefficient vector $\mathbf{A}$ is obtained as:
\begin{equation}
i\hbar \mathbf{S} \dot{\mathbf{A}} = \left[ \mathbf{H} - i\hbar \boldsymbol{\tau} \right] \mathbf{A} \ ,
\label{eq: eom_A}
\end{equation}
where $\mathbf{S}$ is the overlap matrix between configurations,
\begin{equation}
S_{JL} = \bra \Phi_J | \Phi_L \ket = \prod_{\kappa = 1}^M \delta_{j_\kappa l_\kappa} \prod_{\kappa = M + 1}^N S_{j_\kappa l_\kappa}^{(\kappa)} \ ,
\end{equation}
with $S_{jl}^{(\kappa)} = \left\bra g_j^{(\kappa)} \left| g_l^{(\kappa)} \right\ket \right.$;
$\mathbf{H}$ is the Hamiltonian matrix, $H_{JL} = \left\bra \Phi_J \left| \hat{H} \right| \Phi_L \right\ket$, and $\boldsymbol{\tau}$ is the differential overlap matrix with elements
\begin{eqnarray}
\tau_{JL} & = & \bra \Phi_J | \dot{\Phi}_L \ket   \nonumber  \\
 & = & \sum_{\kappa^\prime = 1}^P \left\bra g_{j_{\kappa^\prime}}^{(\kappa^\prime)} \left| \frac{\partial g_{l_{\kappa^\prime}}^{(\kappa^\prime)}}{\partial t} \right\ket \right. \prod_{\kappa = 1}^M \delta_{j_\kappa l_\kappa} \prod_{\substack{\kappa = M + 1 \\ \kappa \neq \kappa^\prime } }^P S_{j_\kappa l_\kappa}^{(\kappa)} \ . \nonumber \\
 & & 
\end{eqnarray}

\item The dynamics of the SPFs of the primary subspace are defined by standard MCTDH equations:
\begin{equation}
i \hbar \dot{\varphi}_j^{(\kappa)} = \left(1 - \mathcal{P}^{(\kappa)} \right) \sum_l \left( {\boldsymbol{\rho}^{(\kappa)}}^{-1} \hat{\mathbf{H}}^{(\kappa)}\right)_{jl} \varphi_l^{(\kappa)} \ ,  \label{eq: eom_SPF}
\end{equation}
where $\mathcal{P}^{(\kappa)}$ is the projector on the space spanned by the SPFs of the $\kappa$-th particle,
\begin{equation}
\mathcal{P}^{(\kappa)} = \sum_j \left| \varphi_j^{(\kappa)} \right\ket \left\bra \varphi_j^{(\kappa)} \right| \ , \ \kappa = 1, ..., M \ .
\end{equation}
The reduced density matrices $\mathbf{\rho}^{(\kappa)}$ and the mean-field operator matrices $\hat{\mathbf{H}}^{(\kappa)}$ are evaluated as
\begin{equation}
\rho_{jl}^{(\kappa)} = \left\bra \psi_j^{(\kappa)} \left| \psi_l^{(\kappa)} \right\ket  \right. \ ,  \label{eq: gmctdh_rho}
\end{equation}
\begin{equation}
\hat{H}_{jl}^{(\kappa)} = \left\bra \psi_j^{(\kappa)} \left| \hat{H} \right| \psi_l^{(\kappa)} \right\ket \ ,
\label{eq: MeanField}
\end{equation}
where the $\psi_j^{(\kappa)}$ are single-hole functions, depending on all coordinates except $\QQ_\kappa$, and defined in such a way that the wavefunction is given as a sum of single-particle functions times the associated single-hole function, i. e.
\begin{eqnarray}
\Psi & = & \sum_j \varphi_j^{(\kappa)} \psi_j^{(\kappa)} \ ,  \ \ \ \ \kappa = 1,...,M \nonumber \\
     & = & \sum_j g_j^{(\kappa)} \psi_j^{(\kappa)} \ ,  \ \ \ \ \kappa = M + 1,...,P \ .
\label{eq: SHF}     
\end{eqnarray}

\item The equations of motions for the GWPs parameters of the secondary subspace have the form
\begin{equation}
i\hbar \sum_{l \beta} C_{j\alpha,l\beta}^{(\kappa)} \dot{\xi}_{l\beta}^{(\kappa)} = Y_{j\alpha}^{(\kappa)} \ , 
\label{eq: eom_GWP}
\end{equation}
where
\begin{equation}
C_{j\alpha,l\beta}^{(\kappa)} = \rho_{jl}^{(\kappa)} \left\bra \frac{\partial g_j^{(\kappa)}}{\partial \xi_{j\alpha}^{(\kappa)}} \left|  1 - \mathcal{P}^{(\kappa)} \left| \frac{\partial g_l^{(\kappa)}}{\partial \xi_{l\beta}^{(\kappa)}} \right\ket \right. \right. \ , 
\end{equation}
\begin{equation}
Y_{j\alpha}^{(\kappa)} = \sum_{l \alpha} \left. \left. \left\bra \frac{\partial g_j^{(\kappa)}}{\partial \xi_{j\alpha}^{(\kappa)} } \right| 
\left( 1 - \mathcal{P}^{(\kappa)}\right) \hat{H}_{jl}^{(\kappa)} \right| g_l^{(\kappa)} \right\ket 
\end{equation}
and the mean-field matrix element $\hat{H}_{jl}^{(\kappa)}$ is defined according to Eqs. (\ref{eq: MeanField}) and (\ref{eq: SHF}). Since the GWPs of a given particle are non-orthogonal, the projector includes the inverse of the overlap matrix,
\begin{equation}
\hspace{-0.5cm} \mathcal{P}^{(\kappa)} = \sum_{jl} \left| g_j^{(\kappa)} \right\ket \left({\mathbf{S}^{(\kappa)}}^{-1}\right)_{jl}  \left\bra g_l^{(\kappa)} \right| \ , \ \kappa = M + 1, ..., P \ .
\end{equation}
\end{itemize}

The G-MCTDH formulation includes the standard MCTDH method as a special case
for $M = P$. The Gaussian-based approximation leads to a significant decrease
of the number of variational parameters, thus allowing a faster integration of
the equations of motion and the quantum mechanical description of many degrees
of freedom with relatively low memory usage.
 However, the
matrices $\mathbf{S}^{(\kappa)}$ and $\mathbf{C}^{(\kappa)}$ can become singular
either when two GWPs are close to each other in phase space or when a small
number of GWPs is populated. In the latter case the configurations $\Phi_J$
which include nearly unoccupied Gaussians $g_{j_\kappa}^{(\kappa)}$ have
small coefficients $A_J$, such that the matrix elements $\rho_{jj}^{(\kappa)}$
are small. In the presence of singularities, the inverse
matrices ${\mathbf{S}^{(\kappa)}}^{-1}$ and ${\mathbf{C}^{(\kappa)}}^{-1}$
must be calculated using regularization procedures (see Sect. \ref{sec: Comp. Det.} and Ref. \onlinecite{BMC99}).

The G-MCTDH method appears particularly suited to model relatively simple molecular systems, which are embedded into a complex environment: The small set of primary particles involved in the \lq system' dynamics are described accurately, whereas the \lq bath' is treated with GWPs. Typical experimental observables are associated with operators acting only on the system degrees of freedom, therefore their simulation requires an accurate evaluation of density matrix $\boldsymbol{\rho}^{(\kappa)}$ only for the primary particles. For the bath coordinates, which are traced out in the evaluation of system properties, the approximate Gaussian description is expected to be adequate. 

The system-environment perspective of the G-MCTDH description is the same as in the Fixed Vibrational Basis/Gaussian Bath (FVB/GB) approach developed by Cina and coworkers\cite{CC07,CCC11,CC14} and successfully applied to the $\mathrm{I_2:Kr}$ system. In the FVB/GB method a different ansatz is used: The energy eigenstates of the system are calculated and the time-dependent wavefunction is obtained by combining each eigenstate with one variationally evolving \lq thawed' Gaussian wave packet (i. e. with time-dependent position and width).

\subsection{Construction of the $\mathrm{I_2 Kr_{18}}$ cluster model}
\label{sec: cluster construction}
The MCTDH and G-MCTDH methods are used to simulate the photo-initiated quantum dynamics of molecular iodine embedded into solid krypton. Wave packet calculations use the same potential energy surfaces (PESs) of the electronic states $X$ and $B$ (see Fig. \ref{fig: PECs}(b)), as in the recent work of Kovac and Cina.\cite{KC17}  The PESs are immediately applicable to any $\mathrm{I_2 Kr}_n$ cluster, since they are force fields defined as a sum of atom-atom pair potentials, and can be easily expressed in terms of the normal modes of the cluster via a simple Gaussian fitting procedure.\cite{KC17} Unfortunately, the resulting Hamiltonian cannot be expressed as a sum-of-products as required by the MCTDH algorithm,\cite{BJWM00} therefore an accurate PES of suitable form for the $\mathrm{I_2:Kr}$ system must be constructed before running quantum dynamical simulations.

To this purpose, a systematic protocol is adopted to construct model PES for an arbitrary number of degrees of freedom. The procedure involves the following steps:
\begin{itemize}
\item A $\mathrm{I_2 Kr}_n$ cluster is obtained from the crystal structure of the solid by including all Kr atoms whose distance from the closest I atom is less or equal to a given threshold $R_1$. Solid Kr forms a face centered cubic lattice and the $\mathrm{I_2}$ molecule replaces two nearest neighbor sites.\citep{AS99} In the present application, $R_1$ is taken equal to $\sqrt{2}$ lattice constants, leading to a $\mathrm{I_2 Kr_{72}}$ cluster. 

\item A smaller cluster enclosed in $\mathrm{I_2 Kr_{72}}$ is identified with the same procedure using a lower distance threshold $R_2 < R_1$. The atoms of the smaller cluster are allowed to be dynamically active, whereas the remaining Kr atoms of $\mathrm{I_2 Kr_{72}}$ are kept frozen in all calculations. In this work, $R_2$ is chosen equal to $\sqrt{2}/2$ lattice constants, so that the small $\mathrm{I_2 Kr_{18}}$ cluster (surrounded by a cage of 54 frozen Kr atoms) is obtained. The equilibrium geometry of $\mathrm{I_2 Kr_{18}}$ has $D_\mathrm{2h}$ symmetry and is highlighted in Fig. \ref{fig: PECs}(a),  where the $\mathrm{Kr}_{54}$ cage is shown as transparent (note that the optimized I--I distance is smaller than $R_2$).

\item The smaller cluster $\mathrm{I_2 Kr_{18}}$, embedded in crystal Kr, is optimized on the electronic ground state $X$. The $60 \times 60$ mass-weighted Hessian matrix $\mathbf{B}$ at the $X$ state minimum is evaluated as
\begin{equation}
 B_{ij} = \frac{1}{\sqrt{M_i M_j}} \left. \frac{\partial^2 V_X}{\partial r_i \partial r_j} \right|_\mathrm{min} \ ,
 \label{eq: Hessian}
\end{equation}
where $r_i$ and $M_i$ are Cartesian coordinates and masses of the cluster atoms and $V_X$ is the $X$ state PES. By diagonalizing $\mathbf{B}$ one gets the normal modes $\tilde{q}_i$ and the corresponding frequencies $\omega_i$, which are strictly positive because the cluster is inserted into the Kr cage. Dimensionless normal coordinates $q_i$ are defined as
\begin{equation}
 q_i = \sqrt{\frac{\omega_i}{\hbar}} \tilde{q}_i \ , \ \ i = 1,...,60 \ ,
\end{equation}
so that the cluster Hamiltonian for a given electronic state $\alpha$ has the form
\begin{equation}
 \hat{H}_\alpha = \sum_{i = 1}^{60} \frac{\hbar \omega_i}{2} p_i^2 + V_\alpha(q_1,...,q_{60}) \ ,
 \label{eq: Ham_60D}
\end{equation}
where $p_i = -i \partial / \partial q_i$ and $q_i = 0$ corresponds to the $X$ state minimum. The vibrational frequencies of $\mathrm{I_2 Kr_{18}}$ were calculated in Ref. \onlinecite{KC17} for the isolated cluster; in the caged cluster used in this work the frequencies of the modes involving vibrations of Kr atoms are larger by a factor $1.1-1.5$.

\item A small subset of normal modes, which are expected to be most dynamically active after the $B \longleftarrow X$ photoexcitation, is identified using classical dynamics simulations. Hamilton's equations are solved using the Hamiltonian of Eq. (\ref{eq: Ham_60D}) for the $B$ state ($\alpha = B$), with initial values of normal coordinates and momenta $\{q_i, p_i\}$ sampled from the Wigner distribution $W_0(q_i,p_i)$ associated with the harmonic ground state of $V_X$,
\begin{equation}
 W_0(q_i,p_i) = \frac{1}{\pi} \exp\left(- q_i^2 - p_i^2 \right) \ .
\end{equation} 
The importance of each mode is established by monitoring the phase space variance of the trajectories with respect to the $X$ minimum. The time-averaged dimensionless quantities
\begin{equation}
 \bar{\Sigma}_i = \left( \frac{1}{T} \int_0^T \left\bra q_i^2 + p_i^2 \right\ket \mathrm{d}t \right) - 1 
\end{equation}
are used to sort the modes in order of their dynamical relevance. In particular, values $\bar{\Sigma}_i \approx 0$ are obtained for the modes whose Wigner distribution remains nearly stationary. 

The time-averaged variances of the 60 modes of the $\mathrm{I_2 Kr_{18}}$
cluster are reported as a bar chart in Fig. \ref{fig: sigma}. The bar of the
mode $q_1$, corresponding to the large amplitude I--I stretch, has a value of
$\bar{\Sigma}_1 = 181$ and is not shown. The bars of the totally symmetric
($a_g$) modes $q_4$, $q_{22}$ and $q_{40}$ clearly stand out, indicating a
significant displacement from the vertical excitation region. Fig. \ref{fig:
sigma}(b) shows that values of $\bar{\Sigma}_i >2$ are found for the modes
$q_{27} \ (a_g)$, $q_{34} \ (a_g)$ and $q_{60} \ (b_{1u})$, which are
therefore also included in the wave packet calculations.

\begin{figure}[t!]
\centering
\includegraphics[scale=0.27]{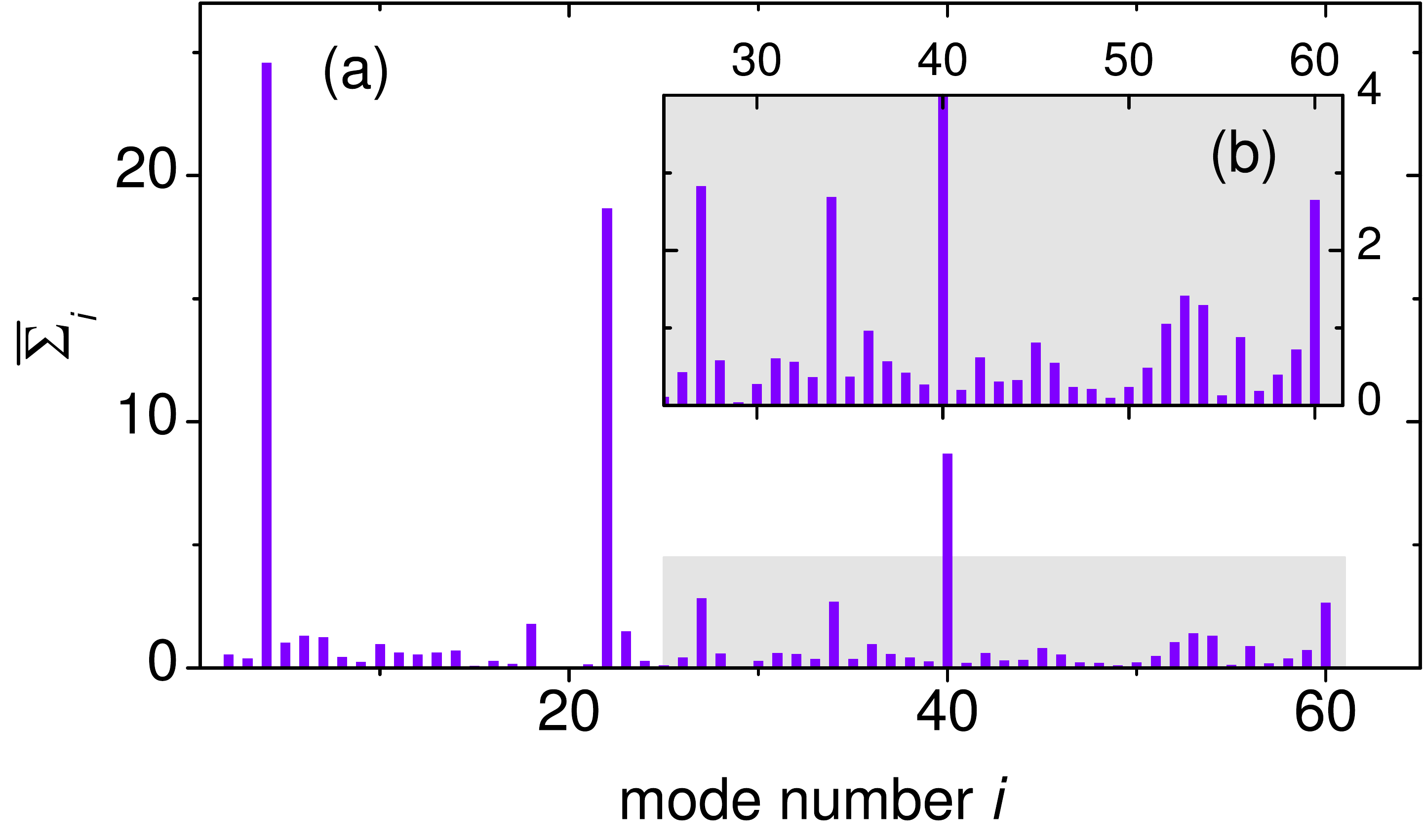}
\caption{(a) Time-integrated phase space variances $\bar{\Sigma}$ of the modes of the $\mathrm{I_2 Kr_{18}}$ cluster embedded in the $\mathrm{Kr}_{54}$ cage, obtained from full-dimensional classical dynamics simulations; the highest variances are found for the modes $q_4$, $q_{22}$ and $q_{40}$. (b) Magnified view of the shaded region; the modes $q_{27}$, $q_{34}$ and $q_{60}$ have $\bar{\Sigma} > 2$.} 
\label{fig: sigma}
\end{figure}

\begin{figure}[b!]
\centering
\includegraphics[scale=0.025]{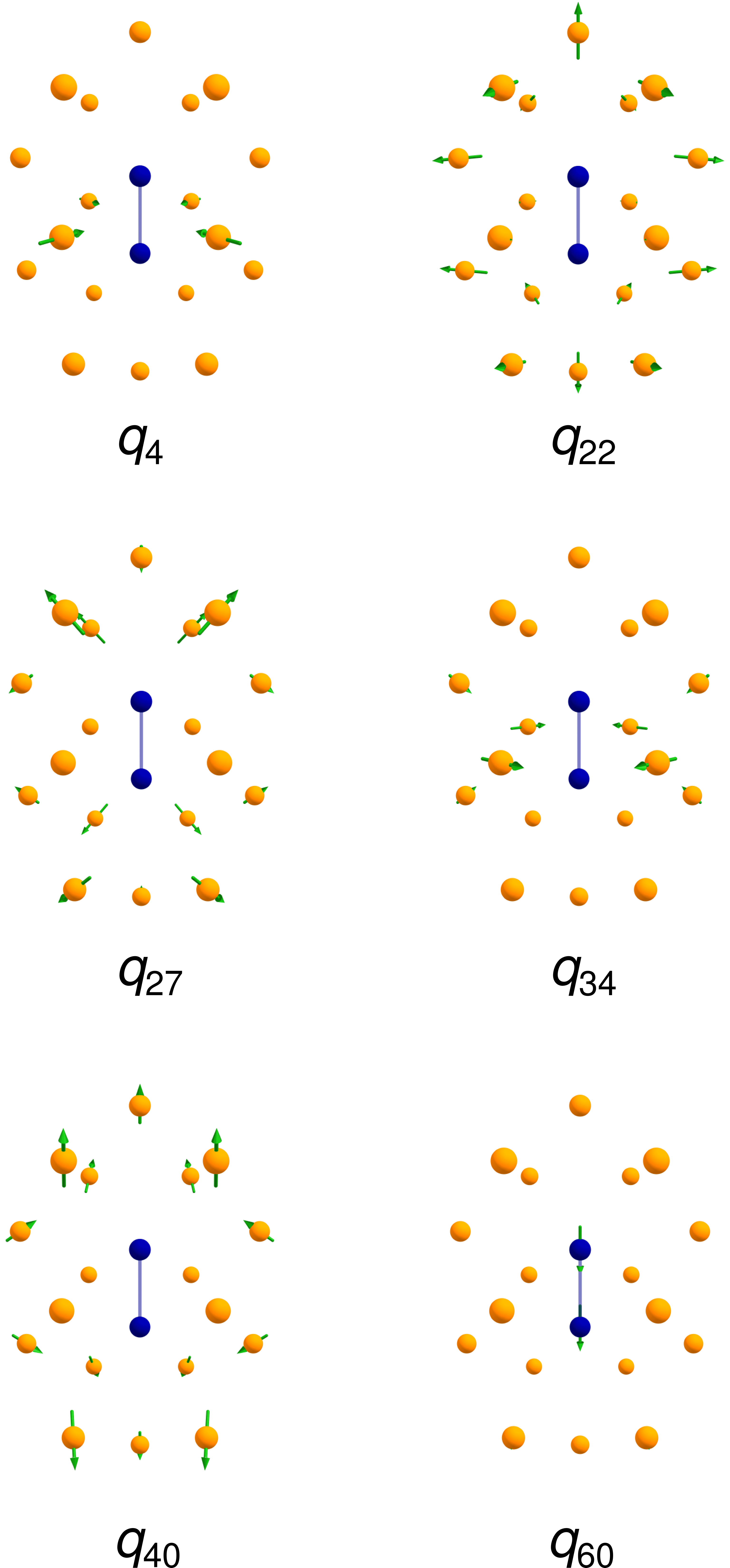}
\caption{The normal modes of the $\mathrm{I_2 Kr_{18}}$ cluster included in quantum dynamical calculations, in addition to the I--I stretching mode $q_1$. } 
\label{fig: ClusterSketch}
\end{figure}

Summarizing, an effective seven-dimensional (7D) Hamiltonian for the caged $\mathrm{I_2 Kr_{18}}$ cluster is constructed; sketches of the six cluster modes which are included in the simulation, besides the I--I stretch $q_1$, are shown in Fig. \ref{fig: ClusterSketch}. Some of the normal modes are reminiscent of the vibrations discussed for the larger scale models of the $\mathrm{I_2:Ar}$ and the $\mathrm{Cl_2:Ar}$ systems,\cite{LZAM95,OA98,FSSK06} and for $\mathrm{Br_2:Ar}$ clusters.\cite{BK07,ABK09} Modes $q_4$ and $q_{34}$ are stretch motions of belt Kr atoms, $q_{22}$ and $q_{40}$ are cage breathing modes, mode $q_{27}$ is the stretch of the eight \lq window' Kr atoms through which the I--I bond can stretch,\cite{GBFKS07} and $q_{60}$ is the pistonlike oscillation of $\mathrm{I_2}$ in the cylindrical host cavity. The frequencies of the seven selected modes evaluated at the $X$ state minimum are reported in Table \ref{tab: normal_modes}. 

\item 7D classical calculations are performed using the Hamiltonian of Eq. (\ref{eq: Ham_60D}) in which the remaining modes are set to zero (i. e. to the $X$ state minimum). The initial phase space sampling is the same as in 60-dimensional simulations. A large number of geometries are sampled from the trajectories (details are given in Sect. \ref{sec: Comp. Det.}); the potential energies of the $X$ and $B$ states at the sampled points are subsequently fitted\cite{note_fitting} to the polynomial functions
\begin{eqnarray}
V(\mathbf{q}) & = & V_0 + \sum_{r = 0}^6 a_{0r} q_1^r  \nonumber \\
 & &   + \sum_{i \neq 1} \sum_{r = 0}^3 a_{ir} q_1^r q_i  + \sum_{i,j \neq 1} \sum_{r = 0}^2 a_{ijr} q_1^r q_i q_j \nonumber \\
 & &    + \sum_{i,j \neq 1} \sum_{r = 0}^1 a_{iijr} q_1^r q_i^2 q_j  + \sum_{i \neq 1} a_{iiii} q_i^4 \ , 
 \label{eq: fitpot}
\end{eqnarray}
which have the desired sum-of-products form and are finally used in wave packet calculations. The coefficients of Eq. (\ref{eq: fitpot}) associated with symmetry-forbidden monomials are set to zero before fitting.  The maximum monomial order in the potentials of Eq. (\ref{eq: fitpot}) is 6; apart for of the terms $a_{05} q_1^5$ and $a_{06} q_1^6$, all the other monomials have maximum order 4. In previous work,\cite{CC14} a polynomial expansion of the potential was also found to be reliable to describe the same system.
\end{itemize}

The advantage of the proposed protocol is its ease of application to different choices of the distance thresholds $R_1$ and $R_2$, which allow one to treat caged $\mathrm{I_2 Kr}_n$ clusters of increasing size.\cite{KC17} Moreover,  a hierarchy of effective Hamiltonians can be generated by the systematic inclusion of additional degrees of freedom associated with lower values of $\bar{\Sigma}_i$. Results obtained for different cluster/cage structures and different sets of dynamically active modes can be eventually extrapolated to the complete crystal. 

From the 7D model of the $\mathrm{I_2 Kr_{18}}$ cluster only a qualitative description of the chromophore-cage interaction can be expected, especially after several I--I vibrational periods. However, one of the main purposes of this paper is to validate the ability of the G-MCTDH approach to describe the nontrivial dissipation and decoherence mechanisms which are operative in a strongly anharmonic van der Waals solid. Allowing direct comparison with exact MCTDH calculations, the 7D model developed in this work is perfectly suitable for this purpose. Moreover, it has been proven in a number of studies that bath dynamics in open quantum systems can be quantitatively reproduced by reduced-dimensionality models which include a small number of \lq effective' bath modes.\cite{CGB05,TBB07,HCB09}

\begin{table}[b!]
\begin{ruledtabular}
\begin{tabular}{cccc}
Mode & $\omega_i \ \mathrm{[cm^{-1}]}$ & Symmetry & Description \\ \hline
$q_1$ & 215.5 & $a_g$ & I--I stretch \\
$q_4$ &  59.6 & $a_g$ & belt atoms breathing \\
$q_{22}$ & 43.2 & $a_g$ & cage breathing \\
$q_{27}$ & 40.2 & $a_g$ & window atoms stretch \\
$q_{34}$ & 35.4 & $a_g$ & belt atoms breathing \\
$q_{40}$ & 32.9 & $a_g$ & cage breathing \\
$q_{60}$ & 17.6 & $b_{1u}$ & pistonlike $\mathrm{I_2}$ motion 
\end{tabular}
\end{ruledtabular}
\caption{Frequencies of the normal modes of the $\mathrm{I_2 Kr_{18}}$ cluster included in quantum dynamical calculations, evaluated at the $X$ state minimum for the cluster embedded into a frozen cage of 54 Kr atoms. For each mode, the irreducible representation of the $D_\mathrm{2h}$ symmetry group is indicated.}
\label{tab: normal_modes}
\end{table}

\section{Computational details}
\label{sec: Comp. Det.}
The optimized geometries of the $\mathrm{I_2 Kr}_n$ clusters and classical dynamics were calculated with in-house developed codes using a Runge-Kutta fourth-order integrator. The force fields for the different electronic states were taken from Ref. \onlinecite{KC17}, and the vertical $B \longleftarrow X$ excitation energy was adjusted to $\mathrm{18500\,cm^{-1}}$.\cite{GBFKS07} In order to select the relevant modes of the $\mathrm{I_2 Kr_{18}}$ cluster, 3000 trajectories were propagated for 5 ps to evaluate the phase space variances $\bar{\Sigma}_i$. For the reduced seven-dimensional system 350 trajectories were calculated for 4 ps; geometries were sampled every 100 fs in order to obtain a data set of 14000 point, which was then used to fit the $X$ and $B$ state PES to the polynomial function of Eq. (\ref{eq: fitpot}). The root mean squared deviation resulting from the fit was $\mathrm{12\,cm^{-1}}$  for both the $X$ and $B$ surfaces.

MCTDH and G-MCTDH calculations were performed using an in-house developed G-MCTDH program. The definitions of the wavefunctions are given in Table \ref{tab: MCTDH details}. The equations of motion (\ref{eq: eom_A}), (\ref{eq: eom_SPF}) and (\ref{eq: eom_GWP}) were integrated for 4\,ps using a variational mean field scheme\cite{BJWM00} with fourth-order Runge-Kutta (for MCTDH) or Adam-Bashfort-Moulton sixth-order (for G-MCTDH) integrators. As usual in MCTDH simulations and in the related variants, the matrices $\mathbf{S}$, $\boldsymbol{\rho}^{(\kappa)}$ and $\mathbf{C}^{(\kappa)}$ were regularized\cite{mctdh-book} to cure singularities, and the integration was performed with an adaptive step size. At each step, the wavefunctions were evaluated for the prescribed integration order $n$ ($\Psi^{[n]}$) and for the order $n+1$ ($\Psi^{[n+1]}$), and the step was repeated with smaller step size if $||\Psi^{[n]} - \Psi^{[n+1]}||> \varepsilon_\mathrm{int}$ ($n = 4$ and $n = 6$ for MCTDH and G-MCTDH, respectively). In the MCTDH calculation a value $\varepsilon_\mathrm{int} = 2\cdot 10^{-7}$ was used and the simulation ran smoothly with step sizes between 0.05\,fs and 0.10\,fs. In G-MCTDH computations it is difficult to get stable integration step sizes for long propagation runs (4\,ps in this case). Because of the singularities due to the non-orthogonality of the Gaussian basis set, decreases of the step size might occur unpredictably during the simulation; in these cases, which occur more often when the number of GWPs is large, the ability to \lq escape the singularity' and to recover larger step sizes depends on the integration accuracy and the regularization parameters. Therefore, such settings were adjusted to the individual G-MCTDH calculations, but the accuracy $\varepsilon_\mathrm{int}$ was always kept between $10^{-7}$ and $6\cdot 10^{-7}$. The optimization of the novel G-MCTDH code and the numerical improvement of the integration stability are currently in progress.

Table \ref{tab: MCTDH details} reports the computation times for the MCTDH and the G-MCTDH calculations. For the reasons just stated, such values cannot be used for a rigorous comparison of performances; however, they might serve as an indication of the speed-up obtained by the Gaussian approximation and provide an estimation of the scalability of the method with increasing number of Gaussians. The accuracy of the G-MCTDH setting I--IV is discussed in Sect. \ref{sec: mctdh vs gmctdh} so that, for the present application, the right compromise between the quality of the G-MCTDH wavefunction and the computation time can be found. 

\begin{table}[h!]
\begin{ruledtabular}
\begin{tabular}{lccccc}
Calculation & Particle & Type & $N$ & $n$ & CPU time \\ \hline
MCTDH & $q_1$    &  DVR &  351  & 18 & 63h 16m\\
      & $(q_4, q_{27})$    &  DVR &  (129,65)  & 20 & \\
      & $(q_{22},q_{34})$ &  DVR &  (113,73)  & 14 & \\
      & $(q_{40},q_{60})$ &  DVR &  (105,89)  & 12 & \\
      & & & & & \\
G-MCTDH  & $q_1$      &  DVR &  351  & 18 & 2h 20m\\
 (I)    & $(q_4,q_{22},q_{40})$ & GWP & -- & 19 & \\
    & $(q_{27},q_{34},q_{60})$ & GWP & -- &  7 & \\
          & & & & & \\
G-MCTDH  & $q_1$      &  DVR & 351  & 18 & 10h 17m \\
 (II)    & $(q_4,q_{22},q_{40})$ & GWP & -- & 33 & \\
    & $(q_{27},q_{34},q_{60})$ & GWP & -- &  7 & \\
          & & & & & \\
G-MCTDH & $q_1$      &  DVR &  351  & 18 & 11h 34, \\
 (III)    & $(q_4,q_{22},q_{40})$ & GWP & -- & 33 & \\
    & $(q_{27},q_{34},q_{60})$ & GWP & -- &  19 &   \\
         & & & & & \\
G-MCTDH & $q_1$      &  DVR &  351  & 18 & 109h 32m \\
 (IV)    & $(q_4,q_{22},q_{40})$ & GWP & -- & 63 & \\
    & $(q_{27},q_{34},q_{60})$ & GWP & -- &  19 &
\end{tabular}
\end{ruledtabular}
\caption{Computational details of MCTDH and G-MCTDH calculations. $N$ and $n$ are respectively the number of primitive DVR grid points and the number of SPFs used for each particle. The type GWP indicates the modes for which a Gaussian representation is used.}
\label{tab: MCTDH details}
\end{table}

\section{MCTDH vs G-MCTDH}
\label{sec: mctdh vs gmctdh}

Quantum dynamical calculations are performed for the 7D model of the caged $\mathrm{I_2 Kr_{18}}$ cluster using the $B$ state Hamiltonian
\begin{equation}
\hat{H}_B = -\sum_i \frac{\hbar \omega_i}{2} \frac{\partial^2}{\partial q_i^2} + V_B(\mathbf{q}) \ ,
\end{equation}
where $\mathbf{q} = (q_1,q_4,q_{22},q_{27},q_{34},q_{40},q_{60}) = (q_1,\mathbf{q}_\mathrm{bath})$ and the potential $V_B$ has the polynomial form of Eq. (\ref{eq: fitpot}). The initial state $\Psi(t = 0)$, depicted in Fig. \ref{fig: PECs}(b), is defined as the harmonic ground state of $V_X$,
\begin{equation}
\Psi(0) = \prod_i \exp\left(- \frac{q_i^2}{2}\right) \ . \label{eq: init state}
\end{equation}

\begin{figure}[t!]
\centering
\includegraphics[scale=0.37]{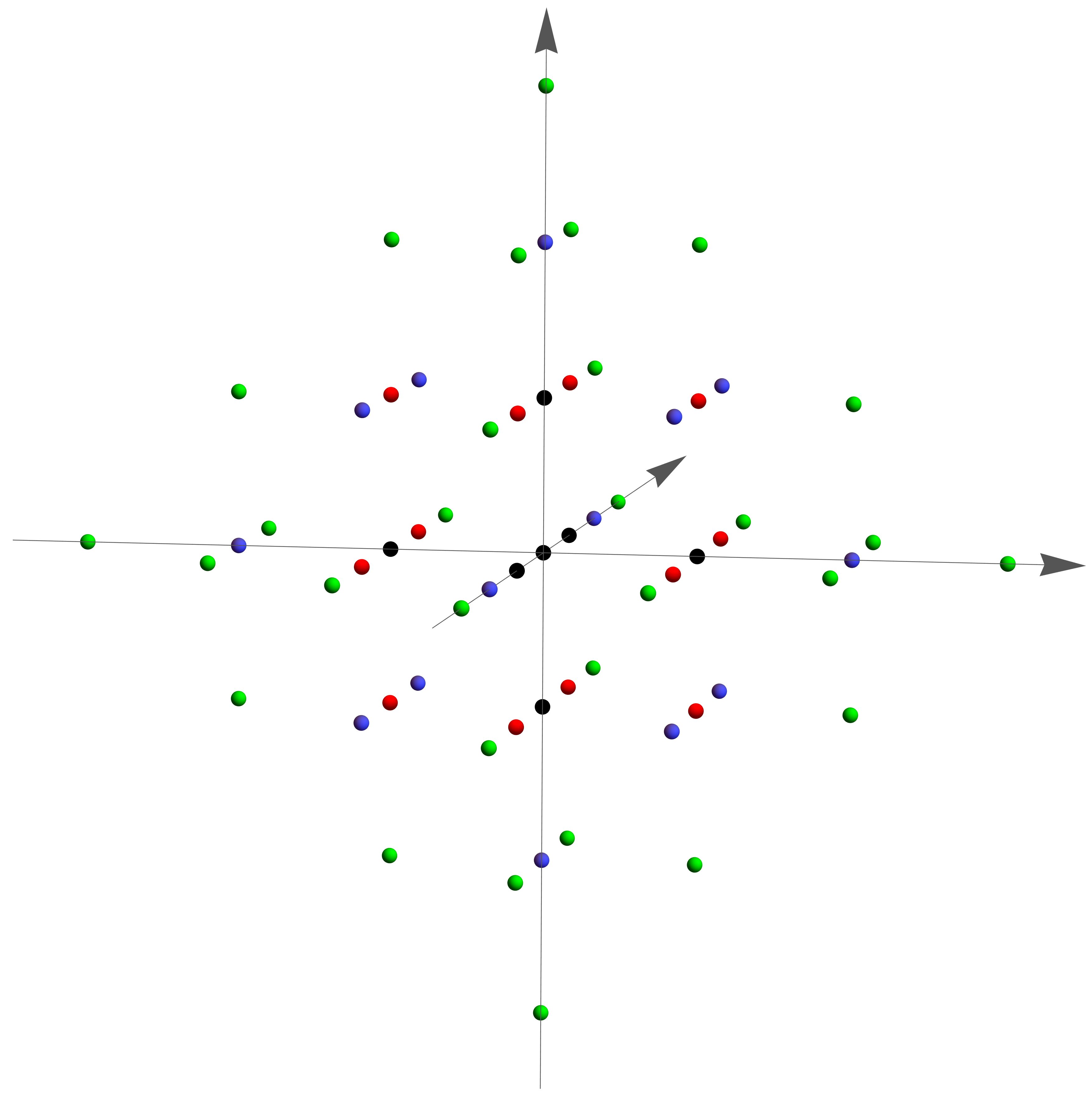}
\caption{Three-dimensional grid of points representing the initial positions of the GWPs used for the secondary modes in the G-MCTDH calculations of Table \ref{tab: MCTDH details}. The three axes represent the three coordinates forming the combined mode. The first shell (black) contains 7 points; the first and second shell (black + red) contain 19 points; the first, second and third shell (black + red + blue) contain 33 points; the four shells (black + red + blue + green) contain 63 points in total.} 
\label{fig: gwp shells}
\end{figure}

Simulations are performed using the MCTDH and G-MCTDH methods, with the MCTDH
results used as an \lq exact' benchmark. Four different settings (I--IV),
reported in Table \ref{tab: MCTDH details}, are used for the G-MCTDH
computations: The difference is the number of SPFs (GWPs) for the Gaussian
particles. In the G-MCTDH wavefunctions, $q_1$ is regarded as primary
coordinate and its SPFs are expressed on a standard sine DVR grid. The
secondary degrees of freedom $\mathbf{q}_\mathrm{bath}$ (i. e. the \lq bath')
are combined into two three-dimensional (3D) particles: The first one includes
the coordinates $(q_4, q_{22}, q_{40})$, which were found to be more active in
classical trajectories (see Sect. \ref{sec: cluster construction}), and is
described with a larger number of GWPs; the second one includes the normal
modes $(q_{27}, q_{34}, q_{60})$ and requires fewer GWPs. The same width is
used for all GWPs. The initially unoccupied GWPs of each 3D mode are initialized following a \lq shell' distribution on a 3D grid,\cite{LRW07} which is the direct product of three one-dimensional (1D) grids. The spacing between the points on the 1D grids are chosen in order that neighboring 1D Gaussians distributed on the grid have an overlap of 0.7. Figure \ref{fig: gwp shells} depicts the 3D grid and highlights the different \lq shells' used in the settings I--IV. In each G-MCTDH calculation the number of GWPs is chosen in order to completely fill a shell.

Below, the quality of the G-MCTDH approximation, as compared to the exact MCTDH wavefunction, is evaluated by checking the accuracy of the autocorrelation function, the Franck-Condon absorption profile, the dissipation rate and the reduced density matrix for the primary coordinate.

\subsection{Autocorrelation function, spectra and dissipation dynamics}
\begin{figure}[b!]
\centering
\includegraphics[scale=0.42]{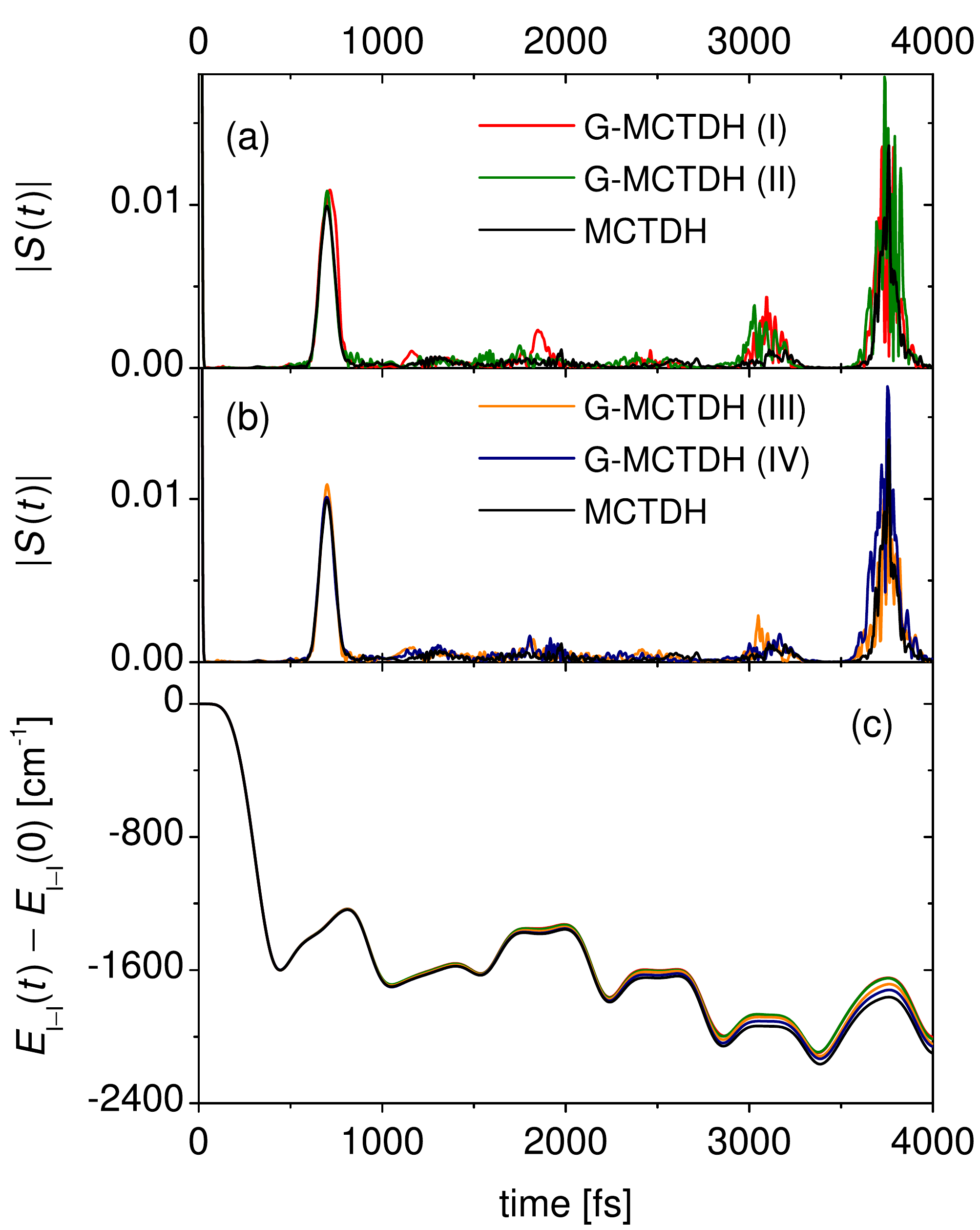}
\caption{(a-b) Absolute value of the autocorrelation function $|S(t)|$ calculated using the MCTDH method (black line) and the G-MCTDH approximation with a different number of Gaussian single-particle functions (red, green, orange and blue lines). (c) Energy loss of the $\mathrm{I_2}$ molecule $E_\mathrm{I-I}(t) - E_\mathrm{I-I}(0)$ calculated using the MCTDH and G-MCTDH methods [same colors as in (a-b)].} 
\label{fig: Auto E_I2}
\end{figure}

The autocorrelation function
\begin{equation}
S(t) = \left\bra \Psi(t) \left| \Psi(0) \right\ket \right. \ , \label{eq: auto}
\end{equation}
is shown in Figs. \ref{fig: Auto E_I2}(a-b) for the first 4 ps of dynamics; the exact MCTDH result is depicted with a black line. In the first $\approx 30$\,fs the function $|S(t)|$ exhibits a falloff from 1 to 0, which is associated with the prompt I--I bond elongation which drives the initial wave packet away from the FC zone. Subsequent tiny recurrence peaks are clearly visible, but their amplitude does not exceed 0.015. This value is one order of magnitude lower than the recurrence amplitudes obtained by Kovac and Cina both for gas-phase $\mathrm{I_2}$ and for the $\mathrm{I_2 Kr_6}$ cluster (which has one totally symmetric bath mode).\cite{KC17} In particular, in the autocorrelation function calculated for $\mathrm{I_2 Kr_6}$, the recurrence associated with the I--I stretching period, is clearly visible as a peak around 400\,fs. In the present 7D model the fastest recurrence, which peaks at 321\,fs, is unnoticeable, because at this time the displaced modes $q_4$, $q_{22}$ and $q_{40}$ are close to their maximum elongation. 
The overlap $\left\bra \Psi(t) \left| \Psi(0) \right\ket \right.$ is therefore strongly affected by the rearrangement of the cage wave packet upon the chromophore photoexcitation.

\begin{figure}[t!]
\centering
\includegraphics[scale=0.22]{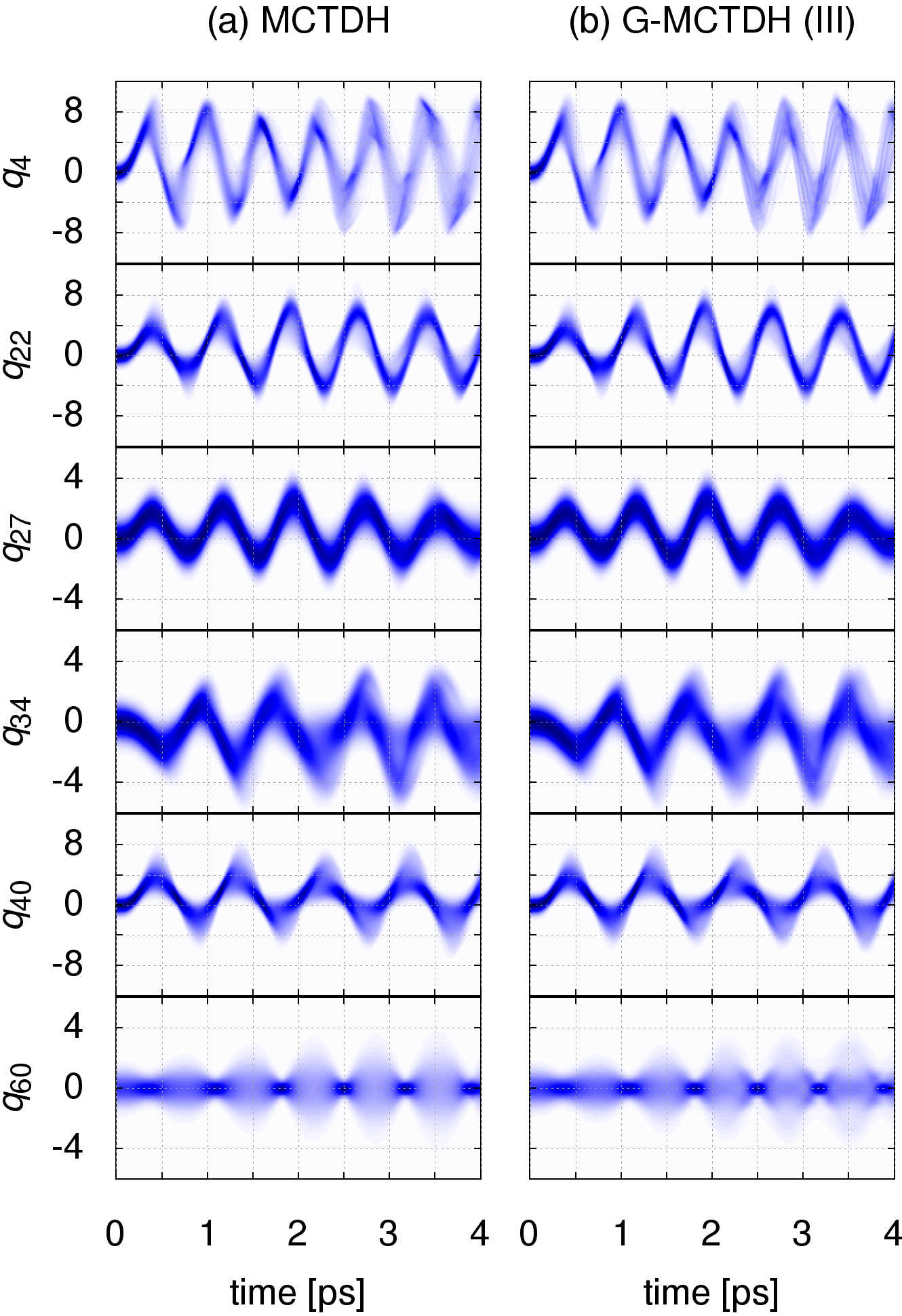}
\caption{One-dimensional reduced probability $\varrho^{(i)}(q_i,t)$ of the bath modes of Fig. \ref{fig: ClusterSketch} as a function of time, calculated using (a) the MCTDH method and (b) the G-MCTDH method with the settings III (see Table \ref{tab: MCTDH details}). }
\label{fig: wave packet}
\end{figure}

In the first 4 ps of dynamics two relatively intense recurrence peaks stand out at 698\,fs and around 3700 -- 3750\,fs. In order to interpret these recurrences it is useful to inspect the time-dependent marginal probability densities
\begin{equation}
\varrho^{(i)}(q_i,t) = \int \left| \Psi(\mathbf{q}) \right|^2 \prod_{j \neq i} \mathrm{d}q_j \ , \label{eq: den1D}
\end{equation}
which are shown, as obtained by the MCTDH calculation, in Fig. \ref{fig: wave packet}(a) for the bath degrees of freedom. 

The cage modes which undergo the largest displacements from the vertical excitation geometries are $q_4$, $q_{22}$ and $q_{40}$, as anticipated by the classical dynamical simulations of Sect. \ref{sec: cluster construction}.
The recurrence time of 698\,fs nearly coincides with the second recursion of the $q_1$ mode. At this time the densities of all cage modes, except for $q_4$, have undergone approximately one oscillation and are close to their initial position, so that $|S(t)|$ becomes (slightly) larger. Similarly, the second partial recurrence of the bath wave packet, occurring at $\approx 3700$\,fs involves mostly the window Kr stretching motion $q_{27}$ and the belt Kr breathing mode $q_{34}$. In the region between 850\,fs and 3500\,fs the amplitude of the autocorrelation function is extremely low and characterized by high-frequency oscillations. Such wiggles are also found in the autocorrelation function of the isolated chromophore\cite{KC17} and are due to the anharmonicity of the I--I interaction potential. Indeed, recurrence times $t_\mathrm{rec}$ are approximately related to the spacing between eigenenergies $\varepsilon_j$ of the $\mathrm{I_2}$ Hamiltonian,
\begin{eqnarray}
\hspace{-0.5cm} \hat{H}_\mathrm{I-I}^{(B)} \chi_j(q_1) & = & \left(- \frac{\hbar \omega_1}{2} \frac{\partial^2}{\partial q_1^2} + V_\mathrm{I-I}^{(B)}(q_1) \right) \chi_j(q_1) \nonumber \\
 &= & \varepsilon_j \chi_j(q_1) \,  \label{eq: HB}
\end{eqnarray}
where $V_\mathrm{I-I}(q_1)$ is obtained from $V(\mathbf{q})$ by setting $q_j = 0$ for all $j \neq 1$. Recursion periods are given as $t_\mathrm{rec} \approx 2n\pi \hbar / |\varepsilon_j - \varepsilon_l|$, with $n$ integer. Since many anharmonic energy levels are initially populated, a number of slightly delayed recursions are operative, so that many closely spaced values of $t_\mathrm{rec}$ emerge in the autocorrelation function. As shown in Sect. \ref{sec: density matrices} the wiggles have a counterpart in the rich nodal structure emerging in the density matrix of the $q_1$ mode after the first $\mathrm{I_2-cage}$ collision.

The autocorrelation functions obtained by the $\mbox{G-MCTDH}$ simulations
I-IV are shown in Fig. \ref{fig: Auto E_I2}(a-b). Even with a small number of
GWPs (calculation I), the $\mbox{G-MCTDH}$ approximation successfully
reproduces the position and amplitudes of the main recurrences. Differences
are however clearly noticeable in the detailed shape of the peaks: For
example, the recurrence at $\approx 3700$\,fs is not fully converged in any of
the G-MCTDH calculations; the behavior of $S(t)$ in the region 850 --
3500\,fs is coarsely described with the settings I and II, but it seems to
converge towards the exact result when two and three GWP shells (settings III
and IV) are included for the bath particles $(q_4, q_{22}, q_{40})$ and
$(q_{27}, q_{34}, q_{60})$, respectively.

One should keep in mind that this is a rather stringent test for the G-MCTDH approach, because the comparison involves an autocorrelation function with very tiny recurrence peaks. It seems therefore unlikely that the inaccuracies of the G-MCTDH results can lead to large errors in the calculation of observables of experimental interest. A more in-depth analysis of the quality of the G-MCTDH approximations and the discrepancies with the MCTDH results is provided by the comparison of the densities of Eq. (\ref{eq: den1D}), which are depicted in Fig. \ref{fig: wave packet}(b) for the bath degrees of freedom, as resulting from the G-MCTDH calculation III. The differences with the MCTDH results are (hardly) noticeable only for the modes $q_4$  and $q_{60}$ at times larger than 2.5\,ps. The G-MCTDH approximation perfectly captures the vibrational periods, oscillation amplitudes, widths and main asymmetries of the bath modes densities. This results in the accurate description of the main cage recursions discussed above.

%\onecolumngrid

\begin{figure*}
\centering
\includegraphics[scale=0.75]{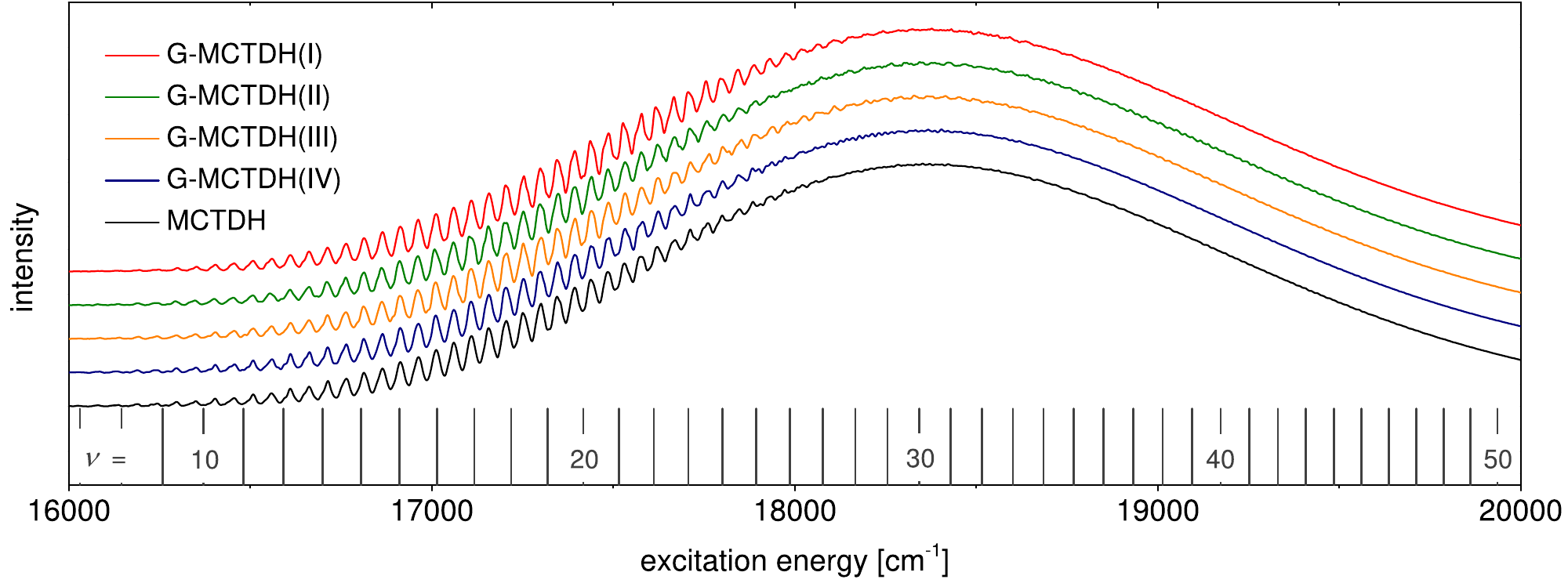}
\caption{Franck-Condon $B \longleftarrow X$ absorption spectra of $\mathrm{I_2 Kr_{18}}$ calculated using the MCTDH method (black line) and the G-MCTDH approximation with a different number of Gaussian single-particle functions (red, green, orange and blue lines).  The combs mark the energy levels of the embedded $\mathrm{I_2}$ and the vibrational quantum number $\nu$ is reported.} 
\label{fig: Spectra}
\end{figure*}
%\twocolumngrid

The Fourier transform of the autocorrelation function yields the linear absorption spectrum,
\begin{equation}
\sigma(\omega) \sim \omega \mathrm{Re} \int_0^T S(t) \mathrm{e}^{i \left(\omega + E_0/\hbar \right) t}  \cos^2\left(\frac{\pi t}{2 T}\right) \mathrm{d}t \, 
\end{equation}
where $E_0$ is the $X$ state (harmonic) energy of $\Psi(0)$ and the autocorrelation function is multiplied by a cosine-squared filtering function ($T$ is the total propagation time). The spectra resulting from the MCTDH and G-MCTDH simulations are shown in Fig. \ref{fig: Spectra}. 

The main features of the MCTDH spectrum are entirely reproduced in the G-MCTDH
calculations. The maximum absorption is near the vertical excitation energy of
$\mathrm{18500\,cm^{-1}}$ and the full width at half-maximum (FWHM) is
$\approx 2000\,\mathrm{cm^{-1}}$. The low-resolution spectral shape is
associated with the initial rapid falloff of the autocorrelation function, and
is due to a long unresolved vibrational progression of the $q_1$ mode. The
energy levels of the embedded iodine, obtained from Eq. (\ref{eq: HB}), are
shown as a comb in Fig. \ref{fig: Spectra} and the vibrational quantum
numbers $\nu$ are indicated; the spectrum covers the range $\nu = 7-60$ and the
maximum is around $\nu = 30-32$, in very good agreement with the analysis of previous work.\cite{OA96,KC17}

  A dense relatively weak sequence of vibrational peaks emerges on top of the
  low energy side of the spectrum. This sequence becomes more pronounced and structured
  between $\mathrm{17100\,cm^{-1}}$ and $\mathrm{17500\,cm^{-1}}$, and
  decreases at higher energies until disappearing at the absorption maximum.
  The spacings between vibrational peaks are in the range
  $\mathrm{45-50\,cm^{-1}}$, which corresponds to the frequencies of the
  totally symmetric bath modes. The vibrational features of the spectrum are
  nicely reproduced in G-MCTDH results, although the structures are too
  pronounced at the center of the spectrum, especially for the settings
  I--III. The peak sequence obtained for the $\mathrm{I_2 Kr_{18}}$ cluster is
  due to long progressions in the totally symmetric cage modes, and is not
  observed in the experimental spectrum on the $\mathrm{I_2:Kr}$
  crystal\cite{KAMAP05} nor in the spectra obtained by semiclassical dynamics
  simulations.\cite{OA96} There are several possible effects that, alone or in
  combination, could lead to a structureless profile for the full solid: (i)
  The presence of additional displaced $a_1$ modes, leading to complete loss
  of recurrences; (ii) finite temperature effects, which result in the
  incoherent average of several spectra; (iii) nonadiabatic or spin-orbit
  driven transitions from the bright $B$ state to neighboring coupled
  states.\cite{ZSA96,OA97,BC97} The validation of the quantum mechanical
  G-MCTDH method to describe the photodynamics of dihalogens in rare gas
  solids is preliminary to the individual study of such effects, which is
  therefore postponed to future work.
 
\begin{figure*}
\centering
\includegraphics[scale=0.23]{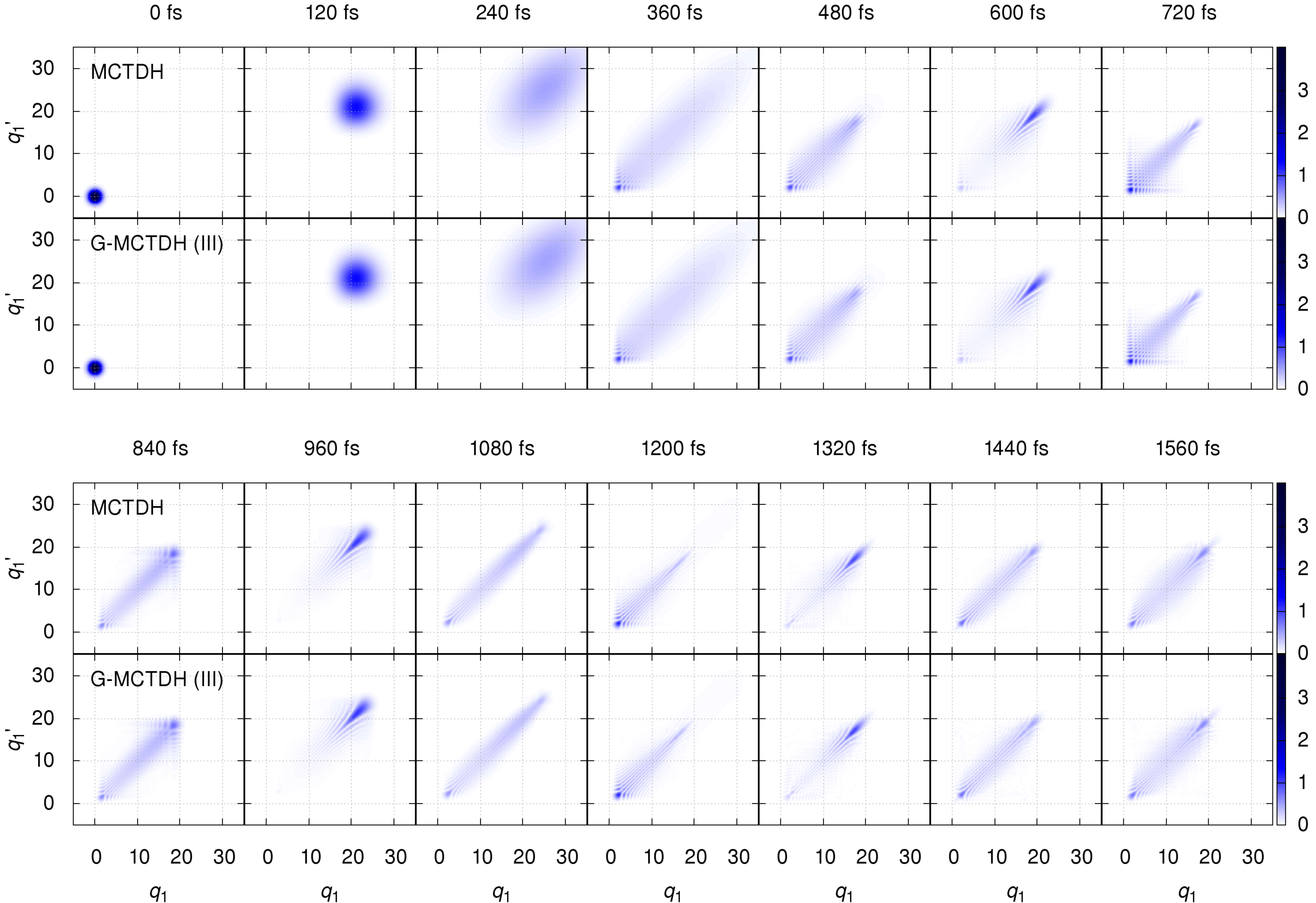}
\caption{Absolute value of the reduced density matrix $|\rho(q_1,q_1^\prime;t)|$ of the I--I stretching mode in the coordinate representation, calculated at different times using the MCTDH and the G-MCTDH (III) methods.} 
\label{fig: rhoQQ}
\end{figure*}

The cage dynamics described above drives dissipation and decoherence processes, which can be therefore monitored via measurements performed on the iodine chromophore. Although the quantitatively correct description of dissipation in the $\mathrm{I_2:Kr}$ system would require several cage phonons, it is worthwhile to analyze how accurately the chromophore-to-environment energy transfer is described by the G-MCTDH approach. The observable of interest is the iodine chromophore energy
\begin{equation}
 E_\mathrm{I-I}(t) = \left\bra \Psi(t) \left| \hat{H}_\mathrm{I-I}^{(B)} + \frac{1}{2}\hat{V}^{(B)}_\mathrm{I-I,bath} \right| \Psi(t) \right\ket \ ,  \label{eq: V_sys_bath}
\end{equation}
where $\frac{1}{2}\hat{V}^{(B)}_\mathrm{I-I,bath}$ represents half of the iodine-environment interaction, which is defined by all terms of Eq. (\ref{eq: fitpot}) which depend on both $q_1$ and $\mathbf{q}_\mathrm{bath}$. With some approximation, the interaction energy is attributed half to the system and half to the bath; as shown in Ref. \onlinecite{NM03} this \lq democratic' splitting is motivated by a virial theorem and becomes exact in the limit of a harmonic bath and a potential $\hat{V}^{(B)}_\mathrm{I-I,bath}$ which is linear in the bath coordinates.
The difference $E_\mathrm{I-I}(t) - E_\mathrm{I-I}(0)$ is plotted as a function of time in Fig. \ref{fig: Auto E_I2}(c). The strongest energy transfer to the Kr cage amounts to $\approx 1600\,\mathrm{cm}^{-1}$ and occurs in the first 450\,fs, i. e. during the first iodine vibrational period. At longer times the dissipation rate decreases, so that only an additional 500\,$\mathrm{cm}^{-1}$ are lost after 4\,ps; however a bath consisting of only six coordinates is expected to provide only a qualitative description of the long time dissipation dynamics of the $\mathrm{I_2:Kr}$ crystal. In the 7D model, the energy loss of iodine is not monotonic because of fluctuations due to the Kr cage motion. In particular, the first oscillation of the dissipation curve occurs at $\approx 800$\,fs; that is the time of the first peak of the autocorrelation function discussed above [compare with Fig. \ref{fig: Auto E_I2}(a)], which was attributed to a recurrence of the bath modes. The energy loss predicted by the G-MCTDH method is in excellent agreement with the exact MCTDH results for the first 2\,ps. At longer times the G-MCTDH curves follow the exact ones but the dissipated energy is slightly underestimated; for the best G-MCTDH (IV) calculation the error is around $\mathrm{30\,cm^{-1}}$ at 3\,ps and $\mathrm{50\,cm^{-1}}$ at 4\,ps. This deviation is less than 3\% of the total amount of dissipated energy and is definitely acceptable in a 4\,ps simulation run. The full convergence of the G-MCTDH results towards the MCTDH prediction is indeed rather slow. The number of bath configurations is more than three times larger in the calculation IV than in the case III, but only a minor improvement is obtained in the description of the bath dynamics.

\subsection{Reduced density matrices in coordinate, Wigner and energy representation}
\label{sec: density matrices}
%\onecolumngrid

%\pagebreak

%\twocolumngrid
The G-MCTDH approach is based on the partition of the degrees of freedom into
primary and secondary coordinates. Since the Gaussian approximation is used
only for the secondary modes, the approach appears suitable to the calculation of
observables associated with operators which involve only the primary modes. As
shown in the companion paper,\cite{PCB18B} in spectroscopic measurements where the
light pulses are short compared to vibrational bath periods the light-matter
interaction can be effectively described with operators independent of the bath coordinates. In order to simulate or interpret measurements of such \lq projected' observables, it is necessary to calculate accurately not the full wavefunction but the reduced density matrix (RDM) for the primary degrees of freedom. 

The accuracy of the G-MCTDH method in reproducing the reduced density matrices of the $\mathrm{I}_2$ chromophore is thus crucial for the simulation of spectroscopic signals. Below, the RDMs calculated with the MCTDH and the G-MCTDH (III) methods are compared for three different representations, which highlight different aspects of the dynamics.

In the coordinate representation, the time-dependent RDM of the mode $q_1$ is defined as
\begin{equation}
\varrho(q_1,q_1^\prime;t) = \int  \Psi(q_1,\mathbf{q}_\mathrm{bath},t) \Psi^*(q_1^\prime,\mathbf{q}_\mathrm{bath},t) \mathrm{d}\mathbf{q}_\mathrm{bath} \ .
\end{equation}
The density matrix is easily obtained during the wave packet propagation as
\begin{equation}
\varrho(q_1,q_1^\prime;t) = \sum_{j,l} \varphi_l^*(q_1^\prime,t) \rho_{lj}(t) \varphi_j(q_1,t)  \ ,  \label{eq: rho_QQ}
\end{equation}
where the $\varphi_j$'s are SPFs and $\rho_{lj}$ is evaluated for the mode $q_1$ according to Eq. (\ref{eq: gmctdh_rho}). The functions $\varrho(q_1,q_1^\prime;t)$ calculated using the MCTDH and G-MCTDH (III) methods are compared in Fig. \ref{fig: rhoQQ} for the first 1560\,fs of dynamics.

%\onecolumngrid 

\begin{figure*} %[h!]
\centering
\includegraphics[scale=0.23]{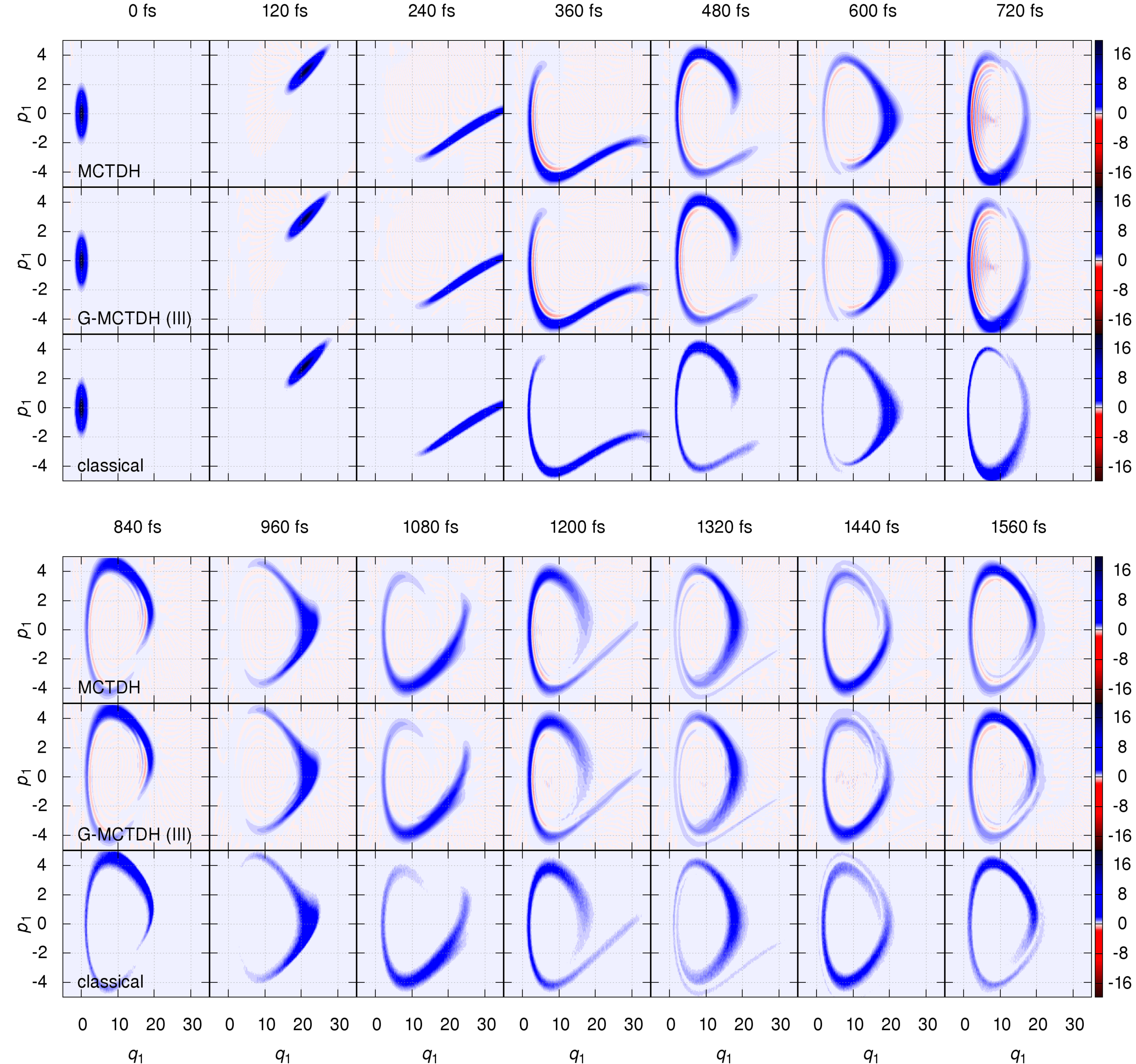}
\caption{Reduced Wigner function $W(q_1,p_1;t)$ of the I--I stretching mode calculated at different times using the MCTDH and the G-MCTDH (III) methods. The phase space distribution $P(q_1,p_1;t)$ obtained by classical dynamics simulations is shown in the bottom row.} 
\label{fig: Wigner}
\end{figure*}
%\pagebreak

%\twocolumngrid

The diagonal $\varrho(q_1,q_1,t)$ is the wave packet density along $q_1$ and the off-diagonal components $\varrho(q_1,q_1^\prime,t)$ (with $q_1 \ne q_1^\prime$) represent spatial coherences between wave packet components located at different values of $q_1$ (i. e. at different I--I bond distances). At time $t = 0\,\mathrm{fs}$ the density matrix has a Gaussian shape and is centered at $q_1 = q_1^\prime = 0$ which corresponds to the initial wave packet of Eq. (\ref{eq: init state}). The way in which the time evolving density matrix changes its shape is a clear manifestation of the anharmonicity of the potential. During the first bond elongation ($\mathrm{0 - 240\,fs}$) $\varrho(q_1,q_1^\prime)$ broadens simultaneously along the diagonal and the anti-diagonal. The components which are spread in different regions of the coordinate space return back to the Franck-Condon zone at different times and with different phases, so that after 300\,fs the density is completely delocalized along the coordinate diagonal. 

Spatially extended coherence between different bond distances is retained for
several iodine oscillations and gives rise to a long-lived nodal structure
along the lines parallel to the axes. The length of such a pattern is a measure
of the extension of spatial coherence. The maximum coherence length is found
-- both at the MCTDH and the G-MCTDH level -- at 720\,fs: For $q_1^\prime
\approx 1$ the coherence $\varrho(q_1,q_1^\prime)$ extends over the range $1 <
q_1 < 15$, which corresponds to a rather broad range of bond distances, $ 2.76\,\mbox{\AA} < R_\mathrm{I-I} < 4.04\,\mbox{\AA}$.
 The presence of long-lived spatial coherences is instrumental to the possibility of creating 
Schr\"{o}dinger cat states, i. e. superpositions of \lq macroscopically different' quantum states,\cite{CL83}  by means of four-wave-mixing optics.\cite{SKFA05,L02}
Interference patterns are observable up to 3.5\,ps and are more prominent close to the vertical excitation zone, where $q_1 \approx 0$ or $q_1^\prime \approx 0$ and the density matrix is less likely to \lq spread'. The G-MCTDH approach nicely reproduces the broadenings, periodic motion and nodal patterns of the exact MCTDH density matrix. The agreement is basically exact for the first ps of dynamics. Small inaccuracies, noticeable at longer times, are mainly found in regions far from the coordinate diagonal and their impact on the expectation values of projected operators is likely to be negligible.
% R = (2.671556 + q1 / 10.951249) a0

%\onecolumngrid

\begin{figure*} %[h!]
\centering
\includegraphics[scale=0.23]{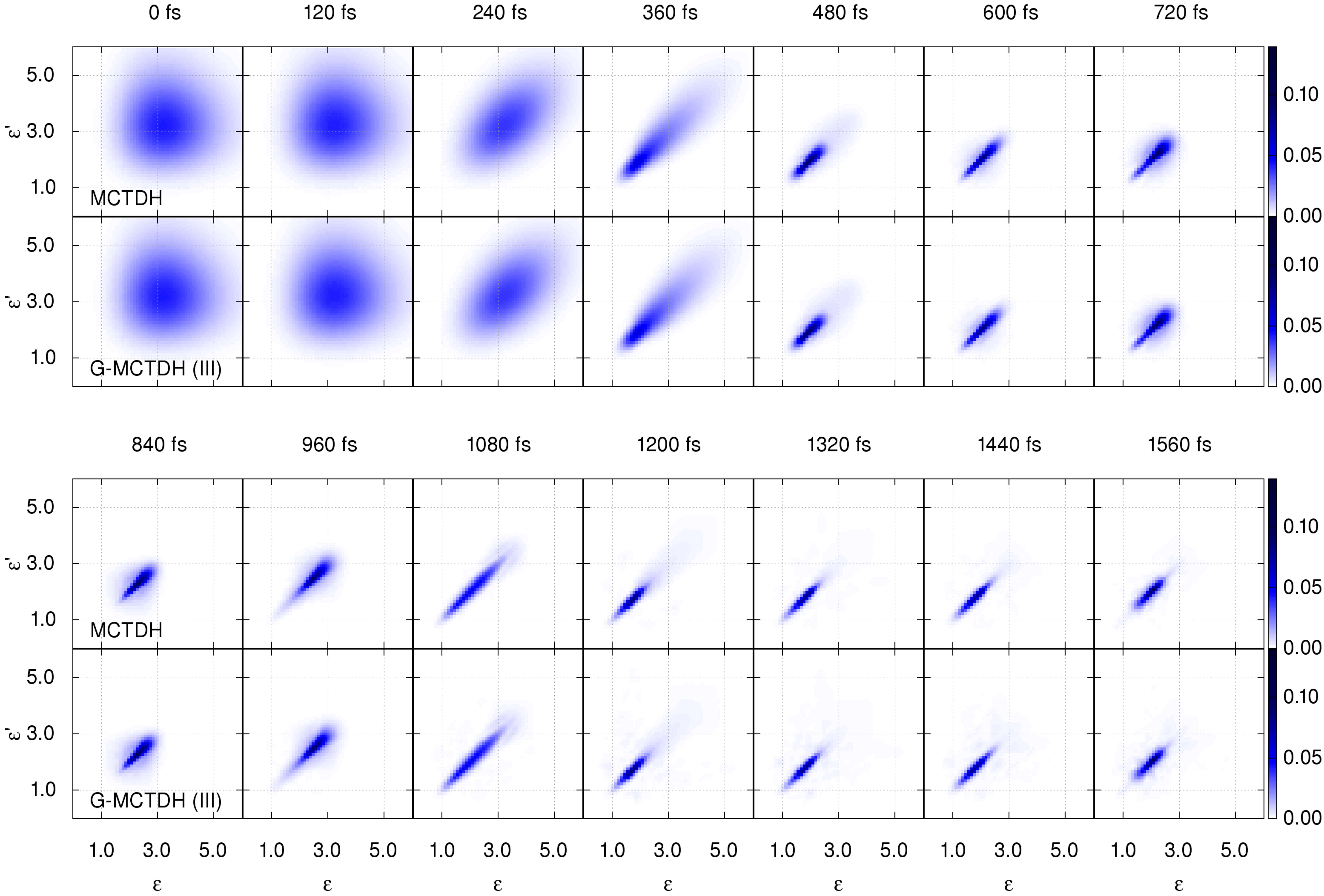}
\caption{Absolute value of the reduced density matrix $|\rho_E(\varepsilon,\varepsilon^\prime;t)|$ of the I--I stretching mode in the representation of the eigenstates of the $B$ state $\mathrm{I_2}$ Hamiltonian, calculated at different times using the MCTDH and the G-MCTDH (III) methods. Energies are shifted by the ground state energy of $\hat{H}_\mathrm{I-I}^{(B)}$ and given in units of $\mathrm{10^3\,cm^{-1}}$. } 
\label{fig: rhoEE}
\end{figure*}
%\pagebreak

%\twocolumngrid

A classical-like picture of the vibrating chromophore is obtained by transforming the reduced density matrix from the coordinate representation to the Wigner picture as
\begin{equation}
W(q_1,p_1;t) = \int \varrho\left(q_1 - \frac{s}{2},q_1 + \frac{s}{2};t \right) \ee^{i s p_1} \mathrm{d}s \ .
\end{equation}
The Wigner functions $W(q_1,p_1;t)$, which represent quasi-probability distributions in the phase space,\cite{L95,quantum_optics_phase_space} are calculated at different times using the MCTDH and the G-MCTDH methods and are depicted in Fig. \ref{fig: Wigner}.
In order to illustrate the quantum mechanical features of the B state dynamics, $N_\mathrm{traj}$ classical trajectories were calculated using the same Hamiltonian, and the classical phase space distribution was obtained as
\begin{eqnarray}
P(q_1,p_1;t) &  \nonumber \\
& \hspace{-2cm} = N_\mathrm{traj}^{-1} \sum_{k = 1}^{N_\mathrm{traj}} \delta\left[q_1 - q_1^{(k)}(t)\right] \delta\left[p_1 - p_1^{(k)}(t)\right] \ , \nonumber \\
& 
\end{eqnarray}
where $\left(q_1^{(k)}(t), p_1^{(k)}(t) \right)$ are classical phase space coordinates for the $k$-th trajectory, and the initial values $\left(q_1^{(k)}(0), p_1^{(k)}(0) \right)$ were sampled from the Gaussian Wigner distribution which mimics the initial wave packet $\Psi(0)$, as described in Sect. \ref{sec: cluster construction}. The time-dependent classical distribution function is shown in the bottom row panels of Fig. \ref{fig: Wigner}.
 At time $t = 0$\,fs $W(q_1,p_1)$ is a two-dimensional positive Gaussian function. During the first bond vibration the distribution evolves semiclassically, tracing a rough orbit in the phase space. After the first bond oscillation ($\mathrm{\approx 350\,fs}$), the spatially extended coherence due to anharmonicity is formed.
  The phase space counterparts of the nodal structure in the coordinate representation are the regions in which the Wigner distribution is negative [they are shown in red in Fig. \ref{fig: Wigner}]. The negativity of $W(q_1,p_1;t)$ is interpreted as a nonclassical effect and its disappearance with increasing time can be used as an indicator of decoherence.\cite{BC99,KZ04} Nonclassical regions emerge most clearly at 720\,fs and for values of $q_1 < 10$ (i. e. bond lengths $R_\mathrm{I-I} < 3.58\,a_0$), which are the time and the coordinate range in which the longest nodal pattern is observed in the density matrix $\varrho(q_1,q_1^\prime)$. 
Regions where the Wigner function is negative are observed for the entire simulation time, up to 4 ps. The motion of the exact MCTDH phase space distribution is extremely well reproduced by he G-MCTDH approximation. The agreement is perfect with regard to the periodicity of the motion, the shape of the orbits and the times of occurrence of large negativities. As with the coordinate picture, the discrepancies occurring after 1\,ps are minor and the G-MCTDH prediction can be considered reliable.

Fig. \ref{fig: Wigner} shows that the global shape and the periods of the Wigner distribution are nicely reproduced by classical simulations. This implies that the bath modes behave nearly classically, and explains why the Gaussian description of the bath converges quickly to the exact MCTDH result. Notice, however, that quantum mechanical interference fringes are absent in a purely classical description; moreover, the Wigner function becomes rather diffuse in the phase space already during the first 300\,fs, so that the convergence of the distribution required a large number of trajectories ($N_\mathrm{traj} = 2\cdot 10^5$). 

In previous work on $\mathrm{I_2}$ in rare gas crystals, classical phase space distributions were used to simulate time-resolved spectroscopic pump-probe experiments.\cite{ZLMA94,LZAM95,BC99} The present results suggest that classical simulations should be suitable especially in this context, because the pump-probe signals monitor the evolution of the density matrix associated with the population (and not the coherence) created by the pump pulse on the excited state.  Quantum signatures in pump-probe spectra are expected only at the times where the negative regions of $W(q_1,p_1,t)$ are more pronounced.  It is worth pointing out, however, that Wigner distributions evaluated in this and in the cited studies refer to an instantaneous electronic excitation; significant differences between the quantum and classical descriptions can be found when the excitation is described by finite duration pulses.\cite{SC99} In test calculations on similar cluster models, with the pulse  interaction Hamiltonian explicitly included (as in Ref. \onlinecite{PCB18B}), more pronounced interference fringes were found in the Wigner functions. In addition, a quantum dynamical treatment is essential for the description of nonlinear coherent spectroscopies, as for example the four-wave-mixing experiments simulated in Paper II,\cite{PCB18B} which monitor the quantum coherence between different wave packets.\cite{KA04,SA11}

A third representation of the reduced density matrix, which is particularly
useful in the interpretation of nonlinear spectroscopies,\cite{PCB18B} is obtained by a projection to the basis of the eigenstates of $\hat{H}^{(B)}_\mathrm{I-I}$ defined in Eq. (\ref{eq: HB}). The time-dependent density matrix is expressed in the energy space as
\begin{eqnarray}
\hspace{-0.5cm} \varrho_E(\varepsilon,\varepsilon^\prime;t) & = & \sum_{jl} \int \mathrm{d}q_1 \int \mathrm{d}q_1^\prime \bra \chi_j| q_1\ket \varrho(q_1,q_1^\prime;t) \bra q_1^\prime | \chi_l\ket  \nonumber \\ 
 & & \hspace{2cm} \times \delta(\varepsilon - \varepsilon_j) \delta(\varepsilon^\prime - \varepsilon_l) \ .  \label{eq: rho_EE}
\end{eqnarray}

The absolute value of the matrices $\varrho_E(\varepsilon,\varepsilon^\prime;t)$ calculated using the MCTDH and G-MCTDH (III) methods are shown in Fig. \ref{fig: rhoEE}; the energies on the axes are shifted by the lowest vibrational level of the $B$ state. The elements on the diagonal of $\varrho_E(\varepsilon,\varepsilon^\prime;t)$ represent the time-dependent populations of the energy eigenstates of the primary system, the off diagonal terms represent coherences between different levels. The energy distribution of the initial wave packet is given by the diagonal $\varrho_E(\varepsilon,\varepsilon;t = 0)$ and has a $\mathrm{FWHM \approx 1900\,cm^{-1}}$. This value is slightly lower than the FHWM of the absorption spectrum (see Sect. \ref{sec: mctdh vs gmctdh}), which accounts also for the small broadening ($\mathrm{\approx 100\,cm^{-1}}$) due to the displaced bath modes. During the first half period of I--I oscillation ($\mathrm{\approx 150\,fs}$) the density matrix preserves its initial shape and, upon the first molecule-cage collision, it starts shrinking mostly along the energy anti-diagonal. This initial \lq shrinkage' represents the ultrafast loss of coherence between eigenstates that are well separated in energy, which occurs simultaneously to the fast strong dissipation (compare with Fig. \ref{fig: Auto E_I2}(c)). After 400\,fs, the density localizes in the low energy region $\varepsilon,\varepsilon^\prime < 3000\,\mathrm{cm}^{-1}$, the dissipation rate decreases and a long lived coherence persists. At 720\,fs the states at $\varepsilon = 2200\,\mathrm{cm}^{-1}$ retain coherence with energy levels which are separated by up to $\mathrm{500-600\,cm^{-1}}$, an energy range which covers 6--7 eigenstates; in the time span 1200--1500\,fs the maximum coherence energy length is around $\mathrm{300\,cm^{-1}}$ (i. e. 3--4 levels). As discussed in Sect. \ref{sec: mctdh vs gmctdh} the dissipation and the related decoherence do not occur monotonically in time, because of the regrowths of vibrational energy induced by recurrences of the bath modes. As illustrated in Fig. \ref{fig: Auto E_I2}(c), one of these recurrences occurs in the time window 950--1100\,fs and is noticeable in the density maps of Fig. \ref{fig: Wigner} at $t = \mathrm{960\,fs}$ and $t = \mathrm{1080\,fs}$, where the density matrix develops a wing extending in the region $\varepsilon,\varepsilon^\prime > 3000\,\mathrm{cm}^{-1}$. The G-MCTDH reproduces the density matrices in the energy picture exactly, even at times longer than 1\,ps.

The agreement with the exact MCTDH results makes the G-MCTDH approach an
effective fully quantum mechanical method to study spectroscopic signals of
chromophores which have few active degrees of freedom and are embedded in a
bath of large dimensionality. As exemplified in the companion paper,\cite{PCB18B} in these cases nonlinear spectroscopies can be often analyzed in terms of the reduced density matrices in the energy picture.\cite{mukamel,CM95,SM03} 
%Given the accuracy of the G-MCTDH method in predicting such quantities, a step forward can be taken and the relatively inexpensive G-MCTDH approach can be used to investigate the theoretical aspects of specific nonlinear spectroscopy experiments.

\section{Conclusion}
\label{sec: Conclusion}
The quantum dynamical simulations performed in this work elucidate the key
aspects of the photodynamics of the iodine chromophore embedded in solid
krypton. Calculations are performed on a seven-dimensional fully anharmonic
Hamiltonian for the $\mathrm{I_2 Kr_{18}}$ cluster, which is derived by
classical dynamics simulations using a standard force field. The most active
degrees of freedom are specific solvent modes, such as the breathing of the
four \lq belt' Kr atoms, which are likely to be operative also in the extended
crystal. The model allows a detailed comparison between the results of a very
accurate MCTDH simulation and the same quantities obtained using the cheaper
G-MCTDH computations. In the G-MCTDH approach the I--I stretch (the system
mode) is treated using standard DVR grids and the Kr cage modes (the bath) are
described using Gaussian wave packets. The Gaussian approximation reproduces
the details of the autocorrelation function, the absorption spectrum, the
$\mathrm{I_2}$ vibrational energy dissipation rate and the reduced densities along the bath
coordinates. Furthermore, all nonclassical features of the reduced density
matrices of the system, e. g. the negative regions of the Wigner distribution
function, are reproduced successfully. The G-MCTDH method can be therefore
recommended for the quantum dynamical studies of chromophores interacting with
solvent modes which behave anharmonically.

Comparison between classical and quantum dynamical simulations shows that a basic description of the dynamics can be obtained by classical trajectories, which do not require a pre-calculated potential energy surface. However, given the large extension of the phase space distribution, a large number of trajectories (of the order of $10^5$) is required to achieve convergence. On the other hand, when the behavior of the bath is nearly classical -- like in this case -- the computationally cheap quantum descriptions based on Gaussian wave packets, like in the G-MCTDH or the FVB/GB methods,\cite{CC07}  converge rapidly to the full quantum result. At a moderate cost, these approaches can thus be used to rigorously account for a number of effects which are difficult to incorporate in classical treatments: The shape of the electronic excitation pulse,\cite{SC99,PCB18B} the entanglement between wave packets created by a sequence of pulses,\cite{SKFA05,SA11} the nonandiabatic transitions,\cite{BC97} and so on.

The reduced dimensionality model of iodine in krypton, necessary to test
validity of the Gaussian approximation, provides a semi-quantitative portrayal
of the dissipation and decoherence dynamics on the $B$ surface. Such picture
will be become more refined after new simulations on larger models, which are
currently being performed using the validated G-MCTDH method and similar
approaches. A multifaceted view of the nontrivial chromophore-cage interactions
is obtained by different representations of the reduced density matrix of the
embedded system. The coordinate representation reveals that quantum mechanical
vibrational coherence is retained over a rather large range of I--I bond
distances ($2.8\,a_0 < R_\mathrm{I-I} < 4.0\,a_0$) and for several vibrational
periods; the negativities of the Wigner function nicely illustrate the
nonclassical features of the dynamics and a slow (longer than 4\,ps)
transition to classicality; the dissipation of the energy initially deposited
in the chromophore is instead easily visualized using the representation of the eigenstates of the caged $\mathrm{I_2}$ chromophore. 

The comparison between different representations shows that the Franck-Condon wave packet retains substantial vibrational coherence even after the initial collisional dissipation ($200-400$\,fs), in which 9-10 vibrational energy quanta are transferred to the krypton cage. In the picosecond time scale the dissipation is slowed down and long lived vibrational coherences are allowed, as a consequence of the lack of resonance between system and bath periods. 

Similar differences between short and long time dynamical regimes are also observed in the two-dimensional model of Kovac and Cina\cite{KC17} discussed above, and in the reduced dimensionality models of Buchholz et al.\cite{BGGSHJ12}. The latter study was performed on a $\mathrm{I_2 Kr_{17}}$ cluster having a geometry different from the crystalline one; a number of wave packet superpositions were constructed on the $B$ state and the purity of the reduced density matrix was calculated as a function of time. Similarly to the dissipation dynamics of Fig. \ref{fig: Auto E_I2}(c), the purity decayed during the first bond elongation; however, in most of the simulations no oscillations due to the bath modes were noticeable, so that the purity remained nearly stationary for 2-3\,ps. This behavior was interpreted as a fast decoherence and indeed no negativities were observed in the reduced Wigner functions of the bath modes. 
On the other hand, coherences between two moving wave packets $\Psi_1$ and $\Psi_2$ are evanescent quantum mechanical features with might be difficult to identify in the observables of the kind $\left\bra \Psi_1 + \Psi_2 \left| \hat{\Omega} \right| \Psi_1 + \Psi_2 \right\ket $, i. e. associated to the full quantum superposition. 

Given the accuracy of the relatively inexpensive G-MCTDH method in predicting reduced density matrices, a step forward can be taken and the approach can be used to investigate the theoretical aspects of coherent nonlinear spectroscopy experiments, which are associated with matrix elements of the kind $\left\bra \Psi_1 \left| \hat{\Omega} \right| \Psi_2 \right\ket$, i. e. with the wave packet coherence. The setup validated in this study is therefore applied in the companion paper\cite{PCB18B} which reports on the
simulation of time-resolved Raman signals revealing the
entanglement and decoherence of \lq Schr\"odinger cat states' as
observed in the experiments by Apkarian and coworkers.\cite{SKFA05,SA11}

\section*{Acknowledgments}
J. A. C. acknowledges the support given by the US-NSF Grant No. CHE 1565680.
D. P. gratefully acknowledges Dr. Pierre Eisenbrandt for beneficial discussions about the development of the G-MCTDH code.

%\bibliography{cluster_GMCTDH}{}

\begin{thebibliography}{78}
\bibitem [{not()}]{note_fitting}%
  \BibitemOpen
  \href@noop {} {}\bibinfo {note} {{\color{black} Alternatively to the proposed
  fit on the dynamical sample, another procedure was tested. Geometries were
  collected performing two- and three-dimensional scans along normal modes, and
  trying to fit the points to similar polynomial functions. In this way,
  however, the fitting sample covered a too large range of energies and it
  included a number of geometries which were far from any of the trajectories
  we found from the initial Wigner sampling. Therefore, a much larger number of
  monomial terms was necessary to converge the fit to reasonable accuracy,
  compared to the dynamical sampling.}}\BibitemShut {Stop}%
\end{thebibliography}


\end{document}